\documentclass{sig-alternate}
\usepackage{tabularx} 
\usepackage{graphicx}
\usepackage{varwidth}
\usepackage{algpseudocode}
\usepackage{algorithm}
\usepackage{amsmath}
\usepackage{tikz}
\usepackage{float}
\usepackage[skip=0pt]{caption}
\usepackage{subcaption}
\usepackage{threeparttable}
\usepackage{multirow}
\usepackage{fancyhdr}
\usepackage[normalem]{ulem}
\usepackage[hyphens]{url}
\usepackage{hyperref}
\usepackage{tabularx} 
\usepackage{graphicx}
\usepackage{varwidth}
\usepackage{algpseudocode}
\usepackage{algorithm}
\usepackage{amsmath}
\usepackage{tikz}
\usepackage{float}
\usepackage[skip=0pt]{caption}
\usepackage{subcaption}
\usepackage{threeparttable}
\usepackage{adjustbox}

\definecolor{myOrange}{rgb}{1,0.753,0}
\definecolor{myBlue}{rgb}{0.1215,0.306,0.4745}
\definecolor{myBlack}{rgb}{0.0,0.0,0.0}

\newcommand*\circledO[1]{\tikz[baseline=(char.base)]{
		\node[shape=circle,draw,inner sep=1.2pt, fill=myOrange] (char) {#1};}}

\newcommand*\circledB[1]{\tikz[baseline=(char.base)]{
		\node[shape=circle,inner sep=1.2pt, fill=myBlue, text=white] (char) {#1};}}
	
\newcommand*\bcircle[1]{\tikz[baseline=(char.base)]{
			\node[shape=circle,inner sep=1.2pt, fill=myBlack, text=white] (char) {#1};}}

\title{Surface Compression Using Dynamic Color Palettes}

\numberofauthors{3} 
\author{
	\alignauthor
	Ayub A. Gubran\\
	\affaddr{University of British Columbia}\\
	\affaddr{Vancouver, BC, Canada}\\
	\email{ayoubg@ece.ubc.ca}
	\alignauthor
	Felix Huang\\
	\affaddr{Carnegie Mellon University}\\
	\affaddr{Pittsburgh, PA, United States}\\
	\email{felixh@andrew.cmu.edu}
	\alignauthor Tor M. Aamodt\\
	\affaddr{University of British Columbia}\\
	\affaddr{Vancouver, BC, Canada}\\
	\email{aamodt@ece.ubc.ca}
	\and  
}

\begin{document}

\maketitle
\begin{abstract}
Off-chip memory traffic is a major source of power and energy consumption on mobile platforms. A large amount of this off-chip traffic is used to manipulate graphics framebuffer surfaces. To cut down the cost of accessing off-chip memory, framebuffer surfaces are compressed to reduce the bandwidth consumed on surface manipulation when rendering or displaying.
 
In this work, we study the compression properties of framebuffer surfaces and highlight the fact that surfaces from different applications have different compression characteristics. We use the results of our analysis to propose a scheme, Dynamic Color Palettes (DCP), which achieves higher compression rates with UI and 2D surfaces. 

DCP is a hardware mechanism for exploiting inter-frame coherence in lossless surface compression; it implements a scheme that dynamically constructs color palettes, which are then used to efficiently compress framebuffer surfaces. To evaluate DCP, we created an extensive set of OpenGL workload traces from 124 Android applications. We found that DCP improves compression rates by 91\% for UI and 20\%  for 2D applications compared to previous proposals~\cite{rasmusson2007exact,nvidiaxwhitepaper}. We also evaluate a hybrid scheme that combines DCP with a generic compression scheme~\cite{rasmusson2007exact}, and found that compression rates improve over previous proposals~\cite{rasmusson2007exact,nvidiaxwhitepaper} by 161\%, 124\% and 83\% for UI, 2D and 3D applications, respectively.
\end{abstract}

\begin{figure*}
	\centering
	\fbox{\includegraphics[width=.5\textwidth]{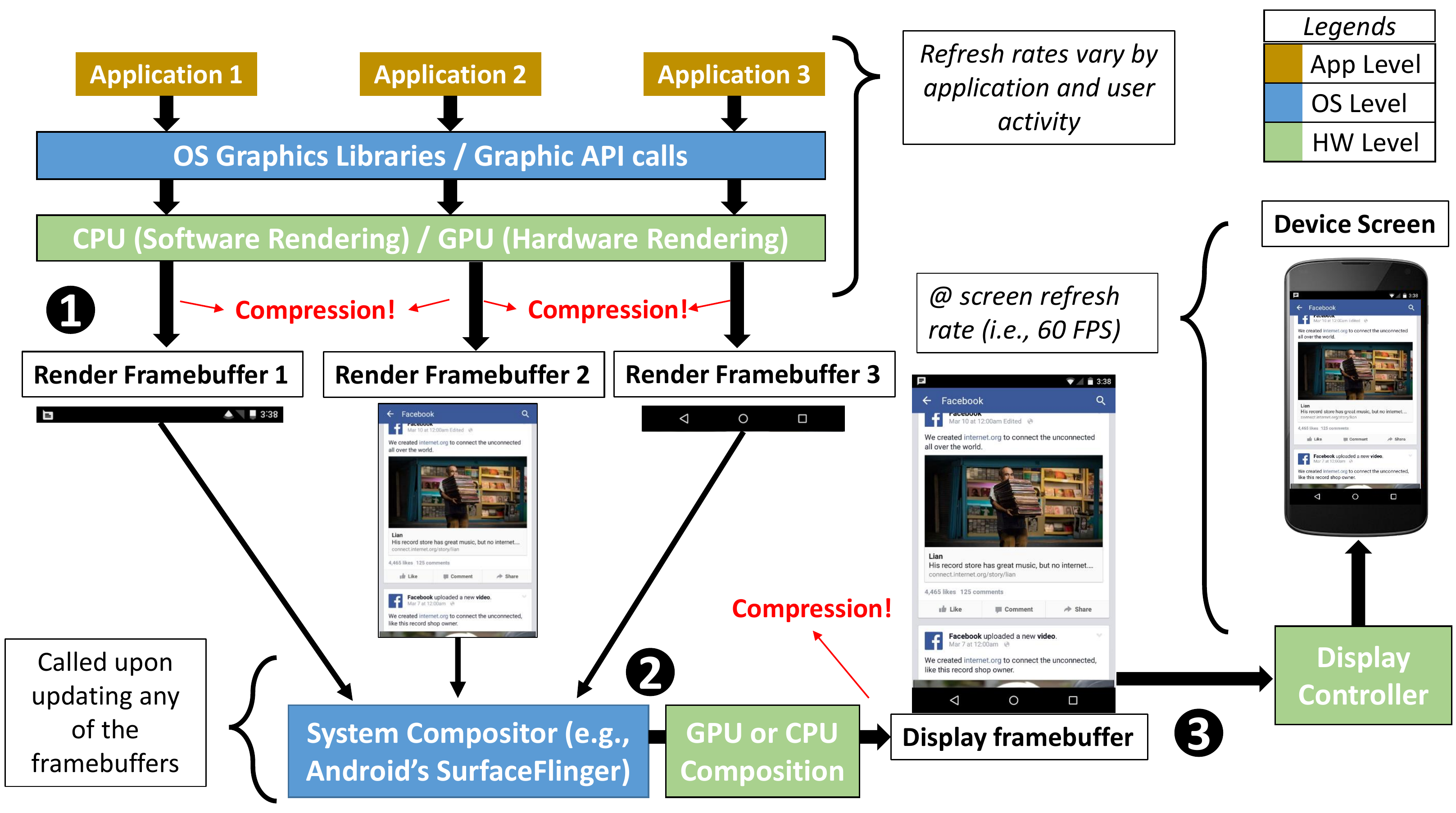}}
	\caption{The life-cycle of a surface from rendering to display. In~\protect\bcircle{1}, applications render to their corresponding framebuffers. In~\protect\bcircle{2}, a compositor combines the surfaces generated by different applications. In~\protect\bcircle{3}, the composited surface is used and displayed on the screen by the display controller.}
	\label{fig:fb_cycle}
\end{figure*}
\section{Introduction}

Off-chip memory traffic, including that of framebuffer surfaces, is one of the major sources of power consumption on mobile systems-on-chip (SoCs). In some cases, the energy consumption to access data on the off-chip memory can dominate that from computations~\cite{olsonmali}. In this work, we study the properties of framebuffer surfaces and propose a set of unique compression techniques to reduce the bandwidth consumed by framebuffer operations.

In graphics rendering, a framebuffer surface is an off-chip memory space that contains pixels generated by the graphics processing unit (GPU) and then used by the display controller to read pixels to the screen. In some cases, the display controller operates on multiple framebuffer surfaces, which are composited to a single surface for screen display. Also GPUs can use framebuffer surfaces as inputs to additional rendering stages, e.g., render to texture and deferred shading; as a result, any given application may utilize one or more framebuffer surfaces.

This work studies a large set of Android workloads to infer the compression properties of framebuffer surfaces generated by mobile UI, 2D and 3D applications. Our study found that framebuffers from different classes of workloads have different compression properties. We exploit these properties to propose an effective palette-based framebuffer compression scheme that focuses on common UI and 2D applications. In addition, we exploit temporal coherence in graphics, where applications exhibit minor changes between frames that can be exploited for compression. 

Using temporal coherence, and by focusing on common uses cases, we propose and evaluate our Dynamic Color Palettes (DCP) technique. DCP uses palette based compression and focuses on reducing the traffic caused by framebuffer operations in UI and 2D applications. To evaluate our compression scheme, we created an extensive set of workloads from 124 Android applications. We show that by combining DCP with other compression techniques~\cite{rasmusson2007exact}, DCP is able to improve compression rates between 83\% and 161\% across UI, 2D and 3D applications.

This paper makes the following contributions:
\begin{enumerate}
	\item Characterizes compression properties of framebuffer surfaces from user-interface (UI) as well as non-UI 2D and 3D applications;
	
	\item Uses characterization results to propose and evaluate dynamic color palettes (DCP), a compression technique that offers higher compression rates for common UI and non-UI 2D applications;
	
	\item Proposes two DCP variations that dynamically choose an optimal palette size based on the frequencies of the values in color palettes;
	
	\item Evaluate our compression schemes using an extensive set of workloads created from the OpenGL traces of 124 Android applications.
	
\end{enumerate}


\section{Background and Related Work}
\label{sec:relatedWork}

\subsection{The life-cycle of a framebuffer surface}

Figure~\ref{fig:fb_cycle} summarizes the life-cycle of a frame surface in contemporary mobile systems (Android Ice Cream Sandwich 4.0 and later~\cite{surfaceflinger}).

Figure~\ref{fig:fb_cycle} shows a typical scenario of drawing multiple surfaces simultaneously from multiple processes: the status bar, Facebook, and the navigation bar. Each process independently renders to its own surface (\bcircle{1}); for example, Facebook renders a new surface when the user scrolls or clicks, while the navigation bar updates the corresponding surface when the user clicks on one of its buttons. 

For display, a system compositor, such as SurfaceFlinger~\cite{surfaceflinger} in Android, combines surfaces from multiple applications before sending them to the screen (\bcircle{2}). 
The compositor actively monitors the surfaces of all applications and when a process updates a surface, the compositor subsequently updates the composited surface. Simultaneously, the display controller hardware continuously reads the composited surface to the screen at 60~frames per second (FPS) or higher (\bcircle{3}). Note that because using the same surface for updates and screen refresh operations can cause artifacts, such as flickering and tearing, double (or triple) buffering is used~\cite{androidArch}. 

The example in Figure~\ref{fig:fb_cycle} shows how a surface can be used and re-used multiple times and this is why it is important to reduce the overhead of framebuffer manipulation through compression.

\subsection{Surface compression techniques}
\label{sec:framebuffer}

Surface compression is used to reduce off-chip memory traffic, which can improve performance and/or reduce energy consumption. Graphics pipeline implementations utilize compression for textures~\cite{strom2005packman}, surfaces~\cite{rasmusson2007exact,akenine2003graphics,antochi2004memory}, depth~\cite{hasselgren2006efficient} and vertex data~\cite{khodakovsky2000progressive}. 

Many of the compression techniques use lossy compression as well on lossless compression. For framebuffer surfaces, lossless compression is used to avoid error accumulation upon reading then re-writing surfaces (as is the case with composition). 

Surface compression differs from texture compression in that both encoding, as well as decoding, are performed in real-time. Also opposite to surface compression schemes, most texture compression algorithms are lossy~\cite{strom2005packman,strom2007etc,nystad2012adaptive}.

Another crucial aspect of surface compression is random accessibility. Techniques like Run-Length Encoding (RLE) are unable to provide such accessibility. However, it important to be able to randomly access a surface when used for sampling (e.g., used as a texture), resizing, or composition.  Compression algorithms have used block-based schemes to enable random access for their simplicity and practicality. Block-based compression mechanisms define preconfigured compression sizes that allow random access to compressed surfaces. Block-based mechanisms have been
used for compressing integer (e.g., RGBA) surfaces~\cite{rasmusson2007exact}, floating-point surfaces~\cite{pool2012lossless,strom2008floating} and
depth buffers~\cite{hasselgren2006efficient,strom2008floating}.

The work by Rasmusson et al.~\cite{rasmusson2007exact} (which we refer to by \emph{RAS}) evaluated several surface compression proposals~\cite{van2006method,molnar2004system,morein2004system} and compared them against their technique. \emph{RAS} is a lossless block-based compression technique for integer buffers that encodes the difference between adjacent pixel values.     
RGB pixels are converted to the $Yc_oc_g$ (luminance-chrominance) format, to increase compression efficiency. 
We compare against \emph{RAS} in this paper since it reports better compression results versus prior work.

We also evaluate our scheme against the compression scheme proposed by Nvidia~\cite{nvidiaxwhitepaper}. In this scheme, for each block going to memory, the algorithm checks if 4$\times$2 pixels in sub-blocks within a block are identical. If so, the block is compressed 1:8. When that is not possible, the algorithm then checks if 2$\times$2 regions have identical colors, if so the block is compressed 1:4, otherwise the block remains uncompressed. This algorithm works well with regions of identical color values, as the case with UI surfaces.

Other compression work includes the work by Danielson~\cite{danielson1998method}, which proposes using a dictionary-based compression
in which the operating system and/or program specify the colors to configure a dictionary.
In contrast to Danielson's work, our work exploits temporal coherence to dynamically construct dictionaries (palettes) avoiding the need for software changes. 

Another work by Shim et al.~\cite{shim2005frame} use a dictionary-based compression mechanism targeted at display buffer compression. 
Shim et al.'s approach compresses surfaces using Huffman coding {\em after} rendering is completed to reduce the bandwidth of display refresh operations. Rendered surfaces are read to construct critical color differences which are used in a second stage to construct a Huffman dictionary. The third stage re-reads the surface buffer and writes out a compressed buffer that is then used for screen refresh operations.
In contrast to previous work, we propose employing temporal coherence to {\em predict} the values for the dictionary, avoiding submitting uncompressed surfaces to memory or requiring additional surface read/write operations. Also we propose an adaptive compression scheme that avoids Huffman coding inefficiencies with probability distributions that are not exact powers of two.

Finally, a body of work has exploited temporal coherence in real-time rendering through inter-frame data reuse. These techniques, in addition to off-line rendering techniques like ray-tracing, are summarized in the survey work of Scherzer et al~\cite{scherzer2012temporal}. Here we propose a different application for temporal coherence by exploiting it for compression.

\subsection{GPU Architectures}
\label{subsec:gpu_archs}
GPU architectures are broadly categorized as either tile-based or immediate-mode architectures.
\label{parag:tilearchs}
Tile-Based architectures aim to save bandwidth by handling all raster operations, like blending and depth testing, using an on-chip buffer. Most mobile GPUs are tile-based architectures (including Qualcomm, Imagination and ARM GPUs). 

In tiled rendering, the screen is divided into render tiles (e.g., 32$\times$32 or 64$\times$64 pixels). For each tile, the GPU renders all primitives that map to that tile using an on-chip buffer before committing that buffer to the off-chip memory. As a result, what is being compressed and sent to the off-chip memory is the final surface value of each tile.

On the other hand, immediate-mode architectures render primitives in their drawing order. They avoid the overhead of the tiling process, but the values sent to the off-chip memory will contain intermediate surface values. In the case of overdrawing (i.e., when a pixel location is covered by more than one primitive), the same memory location will be written to multiple times. When a compression scheme is deployed in an immediate-mode GPU, it will compress the values sent to the off-chip memory as rendering progresses; this means compressing blocks from the GPU's LLC instead of an on-chip buffer.

As tile-based architectures are the dominant choice for mobile GPUs, going forward, we assume a tile-based architecture when evaluating surfaces for compression.

\subsection{Mobile Use Patterns}
In this work, we focus on developing an effective scheme for compressing UI and 2D framebuffer surfaces. The reason for this choice is that studies found users to spend 70\% of their time running UI applications~\cite{flurry2013,nielsen2011android}, where over half of the time is spent on web browsing, messaging and social media alone. Whereas, games of all types, 2D, and 3D, account for only 30\% of the usage time. Thus, we saw the opportunity of designing an effective scheme that targets such common use cases. In Section~\ref{subsec:hybrid}, we show how our DCP scheme can be combined with other generic compression algorithms to provide a comprehensive compression solution for all use-cases. 

\section{Temporal Coherence in Mobile Graphics}

\label{sec:temporalcoh}

Temporal coherence is the property of inter-frame similarity~\cite{scherzer2012temporal}; this means that in a sequence of frames, content only  gradually changes from one frame to the next. To quantify temporal coherence, we use two measurements: \emph{Color change} and \emph{Pixel change}.

\emph{Color change} is the total difference in pixel color frequencies between two frames regardless of the locations of the pixels. On the other hand, \emph{Pixel change} is the total number of pixels that change color between frames, which is measured by counting the number of pixel locations that differ in value between two frames. \emph{Color change} estimates how similar two frames are, only with regard to color frequency. While \emph{Pixel change} captures the movement of content on a surface.

\begin{figure}
	\centering
	\includegraphics[width=.5\textwidth]{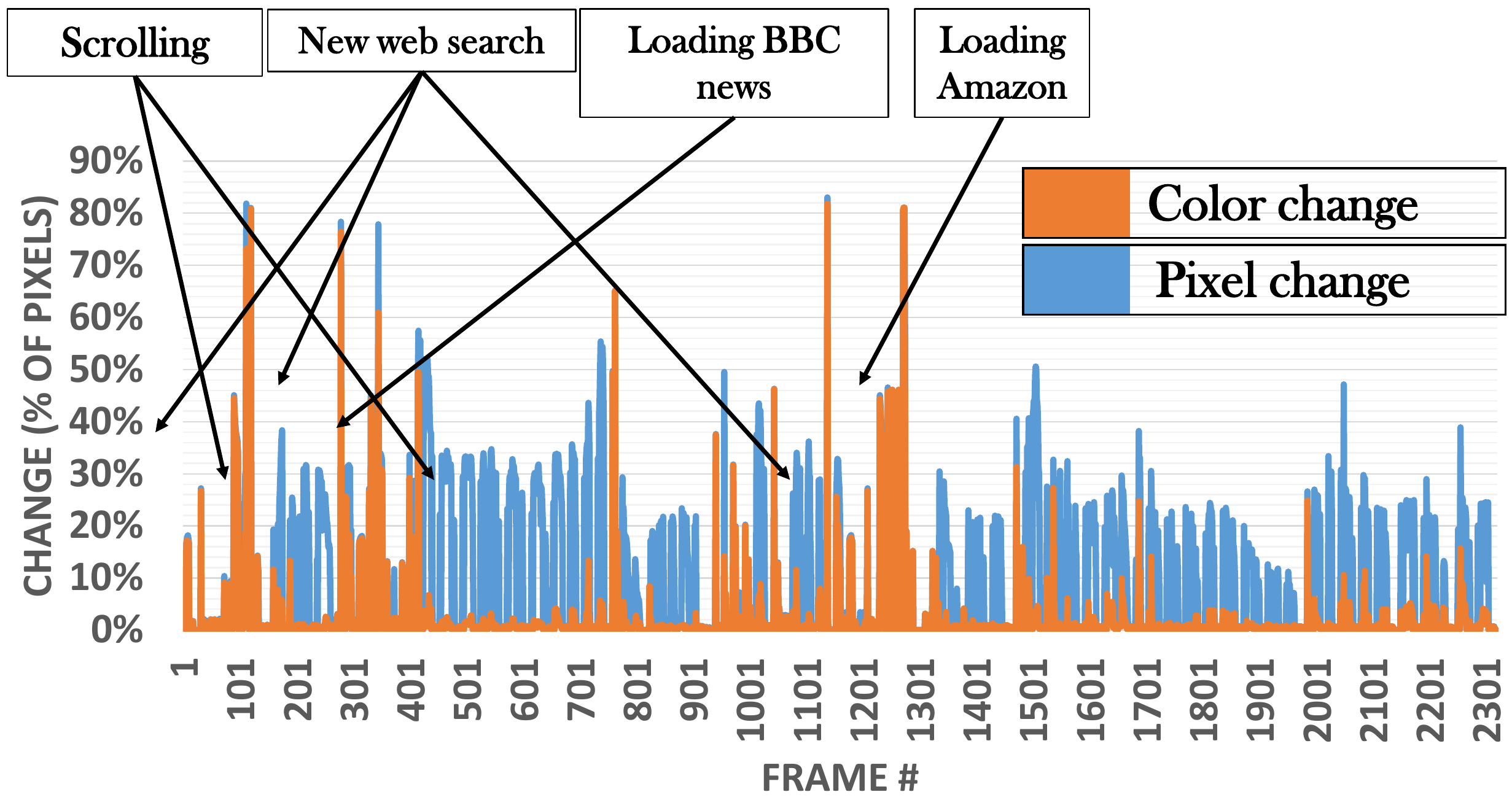}
	\caption{\emph{Pixel change} and \emph{Color change} in \textbf{Google Chrome}. In most cases, pixel change is higher than color change; this means that content is moving but not changing most of the time. Our compression scheme takes advantage of lower color change between frames to predict compression palettes.}
	\label{fig:chrome_change}
\end{figure}

To illustrate color and pixel change, we use an example from the \textbf{Google Chrome} browser in Figure~\ref{fig:chrome_change}. The example shows pixel and color change for different events. Notice that in some cases, when a new content is displayed, both pixel and color change values are high (e.g., \textit{new web search}). In most cases, however, pixel change is always higher than color change; this means that in many cases the content is moving but not changing, as the case with \textit{BBC news} in Figure~\ref{fig:chrome_change}. 

Looking at a range of mobile workloads, we found that temporal color coherence is reflected by low \emph{Color change} values, especially in UI applications. We analyzed a set of nine Android UI applications and games (UI: Twitter; Facebook; Chrome; and Android Home Screen, and 3D: Fruit Ninja; Need 4 Speed; Gunship 2; and Temple Run 2). In 3D applications, color and pixel change rates are 15.7\% and 65\%, respectively. On the other hand, UI applications has rates of 3.3\% and 14.5\% for color and pixel change values, respectively. These numbers show that 2D and UI applications exhibit higher temporal \textit{color} coherence relative to 3D applications.


\begin{figure*}
	\centering
	\begin{tabular}{@{}c|c@{}}
		\includegraphics[width=.25\textheight]{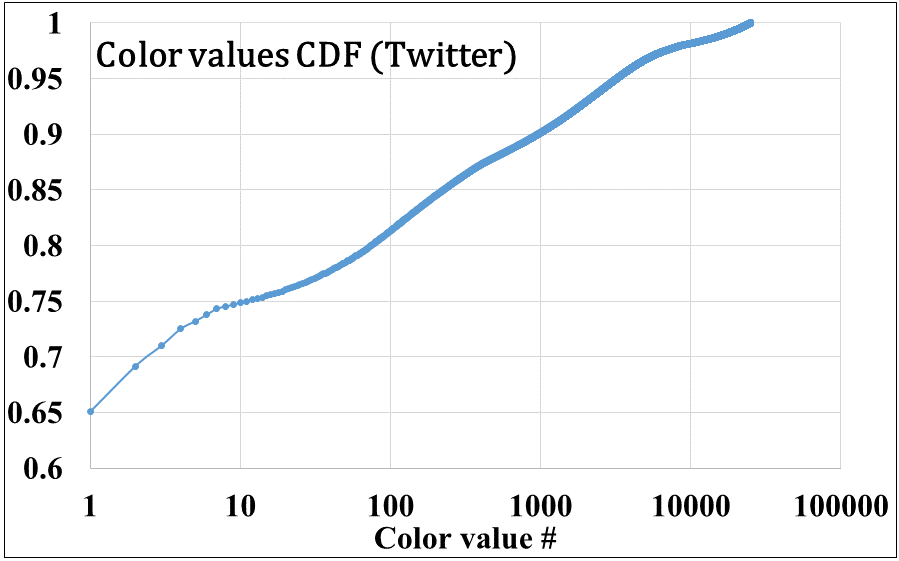}
		\includegraphics[height=.18\textheight]{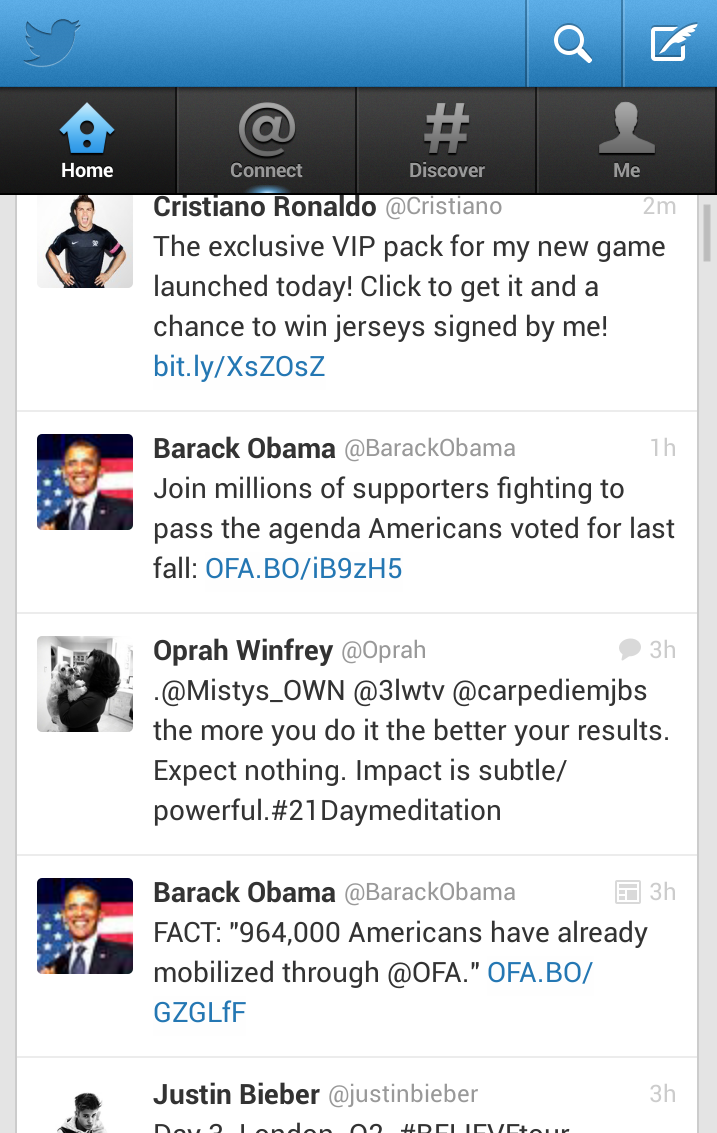} &
		\includegraphics[height=.18\textheight]{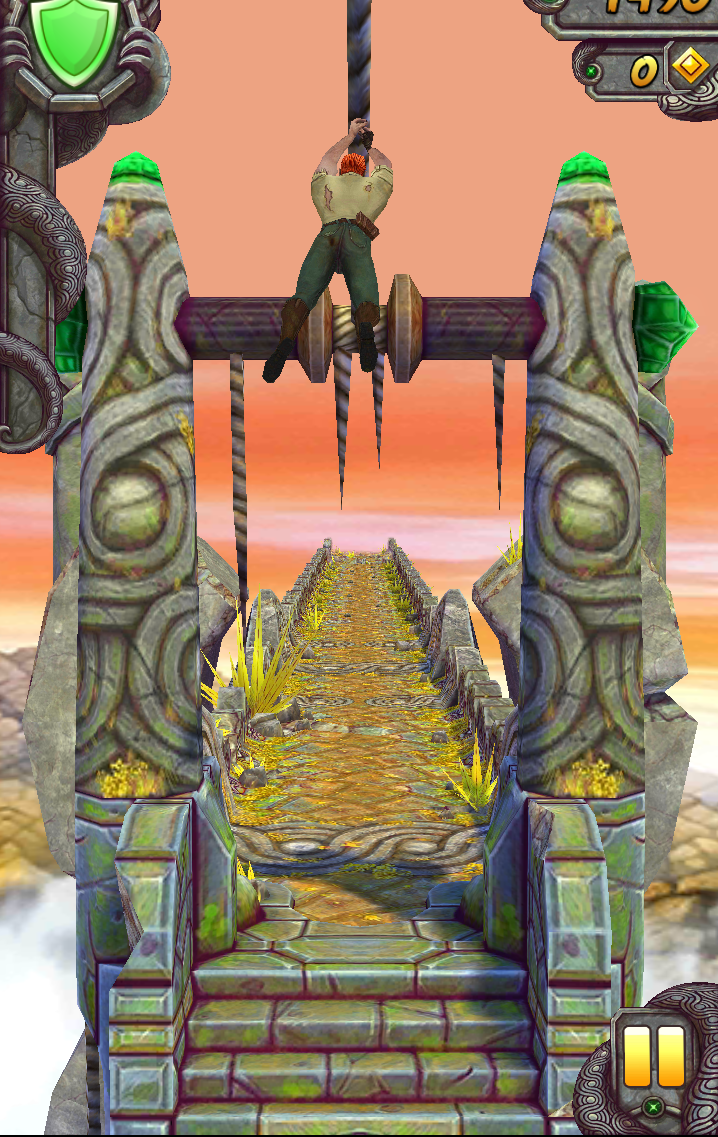}
		\includegraphics[width=.25\textheight]{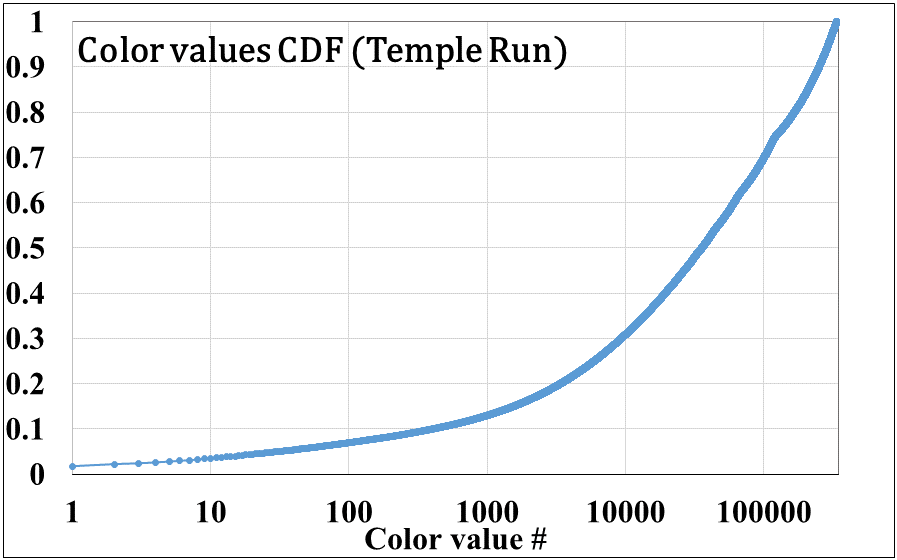}
		\\ 
		\textbf{(a)} Twitter & \textbf{(b)} Temple Run 2
	\end{tabular}
	\caption{The cumulative distribution function (CDF) of unique color values in UI (Twitter) and 3D (Temple Run 2) Android applications.}
	\label{fig:d3colors}
\end{figure*}

In addition to higher temporal coherence, we found that UI applications tend to use fewer colors. Figure~\ref{fig:d3colors} demonstrates how a small number of frequent pixel color values dominates a typical UI application compared to a 3D one. Figure~\ref{fig:d3colors} shows the cumulative distribution function of colors used in Twitter UI (a), compared to a 3D game, Temple Run 2 (b). In (a), the top 100 most common color values cover over 80\% of the frame's surface, while coverage is 10\% for Temple Run 2. Measuring compressibility with Shannon entropy, we found that Twitter has an entropy of 4.5 bits per pixel, while it is 14 bits per pixel for Temple Run, indicating higher compressibility for Twitter. 

The next section shows how to take advantage of temporal coherence and the color characteristics of UI applications to design a dynamic color palette scheme for compressing UI and 2D surfaces.


\section{Dynamic Color Palettes (DCP) Compression}
\label{sec:dcp}

\begin{figure}
	\centering
	\includegraphics[width=.5\textwidth]{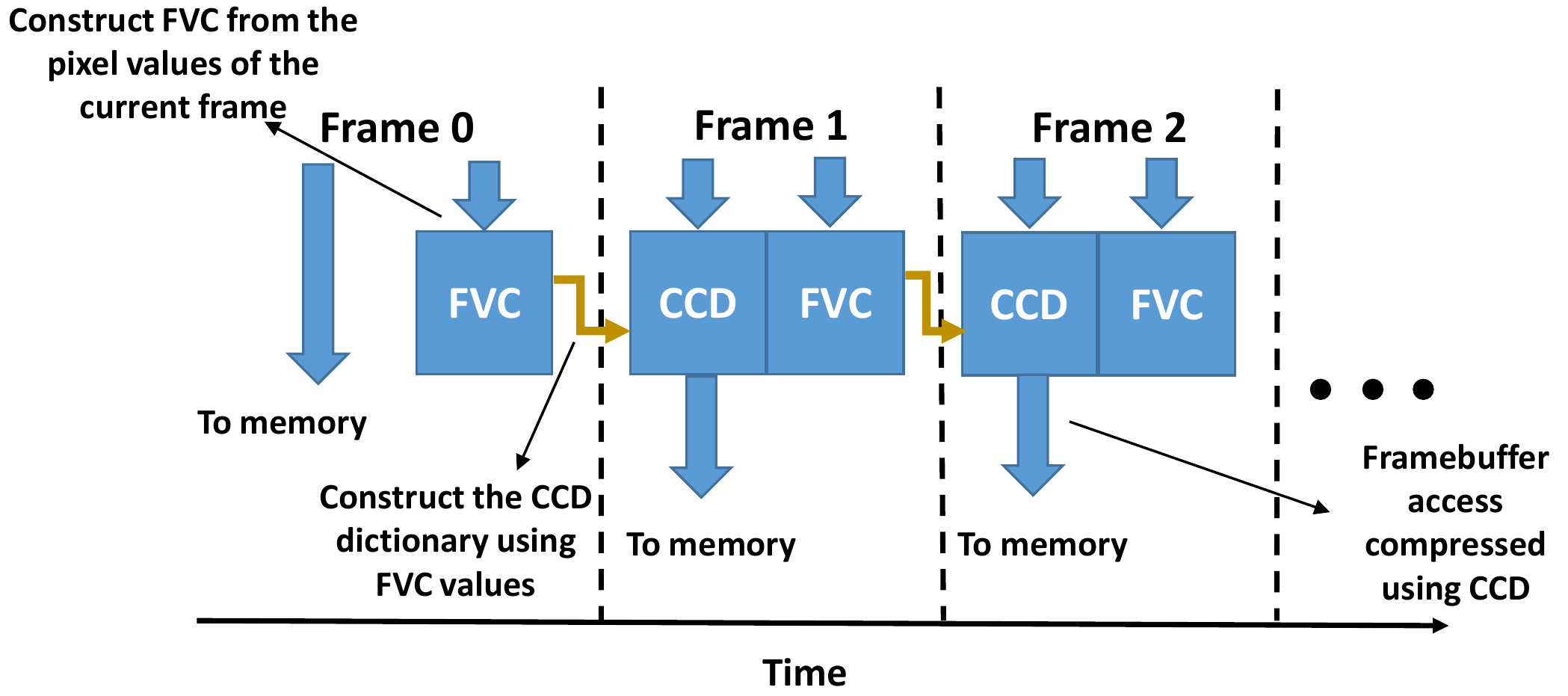}
	\caption{Using DCP across frames.}
	\label{fig:dcp_steps}
\end{figure}

\begin{figure}
	\centering
	\includegraphics[width=0.5\textwidth]{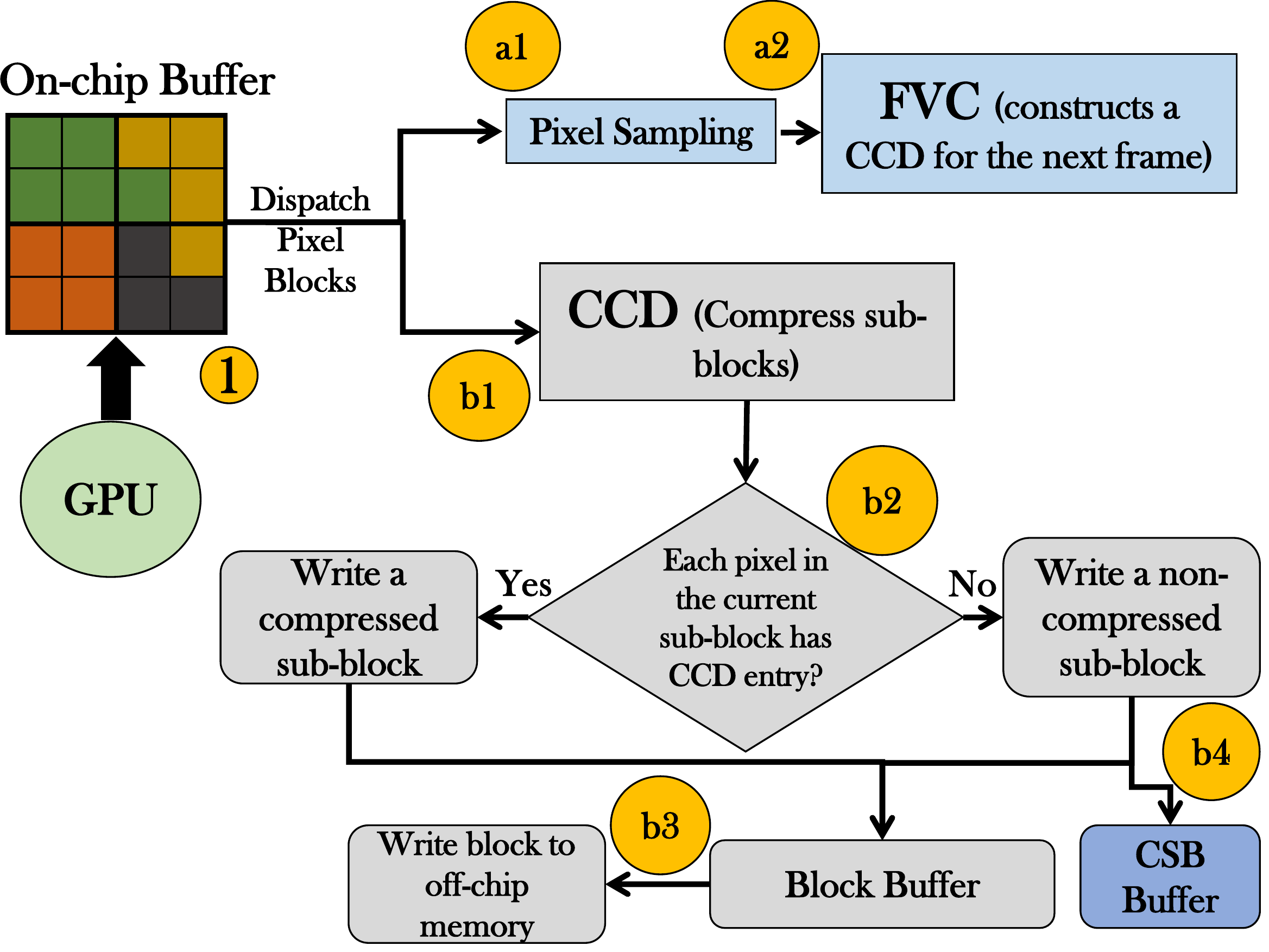}
	\caption{DCP stages.}
	\label{fig:dcp_stages}
\end{figure}

DCP is a technique to exploit graphics temporal color coherence for framebuffer compression. For each frame, DCP carries two operations in parallel: color frequency collection and framebuffer compression. For color frequency collection, DCP tracks the most frequently used colors as the rendering of a frame progresses; meanwhile, DCP works on compressing the pixels in the frame with a palette constructed using the frequency information of the previous frame. 

DCP has two main advantages over previous dictionary-based techniques~\cite{danielson1998method,shim2005frame}. First, it employs sampling to exploit temporal coherence to predict future dictionary values on-the-fly, alleviating the need for software hints or a multi-stage dictionary update process. This allows DCP to compress intermediate surfaces (i.e., application surfaces) as well as the framebuffer surface used by the display unit.  Second, as we will show later, DCP maximizes compression using adaptive dictionary sizing, which puts to use the color frequency data collected for each frame.

DCP relies on two structures (shown in Figure~\ref{fig:dcp_steps}), the Frequent Values Collector (FVC) for color frequency collection, and the Common Colors Dictionary (CCD) for compressing new pixels. 
The FVC identifies most commonly occurring colors, while the CCD encodes the most frequent colors as identified by the FVC from the previous frame. As shown in Figure~\ref{fig:dcp_steps}, each frame the FVC collects color frequency information that are then used to construct the CCD of the next frame. 

\subsection{DCP workflow}
Figure~\ref{fig:dcp_stages} shows DCP workflow. In \circledO{1}, the GPU commits tiles to the off-chip memory in multiple batches, i.e., blocks of spatially adjacent pixels~\cite{lpddr2jedec,lpddr3jedec,lpddr4jedec}. For the example in Figure~\ref{fig:dcp_stages}, we use a block size of 4$\times$4 pixels and a sub-block size of 2$\times$2. Pixels in each block are sent to the FVC~\circledO{a1} and the CCD~\circledO{b1}.

In \circledO{a1}, the FVC uses pixel values in each block to update the common color frequencies of the current frame (more details on that in the next Section).

In \circledO{b1}, the CCD compresses pixel blocks in batches of sub-blocks. In~\circledO{b2}, if all pixel values in a sub-block have an entry in the CCD, the sub-block is determined to be compressible. Each color value in a compressible sub-block is represented using $log_2($CCD size$)$ bits, e.g., 6 bits per pixel for a CCD with 64 entries. If one of the pixel value in a sub-block does not have a CCD entry, then the whole sub-block remains uncompressed. Compressed and non-compressed sub-blocks are buffered and once a full block is processed, it is then written to the off-chip memory~\circledO{b3}. 

Like other block-based compression schemes~\cite{rasmusson2007exact,kulshrestha2010selecting}, DCP uses an a metadata compression status buffer (CSB) that contains a compression status bit for each sub-block. Upon compressing a sub-block, the corresponding entry in the CSB is set~\circledO{b4}, and upon reading a compressed surface, the CSB is consulted to determine how much data should be fetched from memory. 


\subsection{The Frequent Values Collector (FVC)}
\label{subsec:fvc}
FVC is a relatively small--e.g., 16 to 128 entries--associative memory structure. The FVC stores a set of pixel values and their corresponding frequencies as value-frequency pairs. For each pixel access, the FVC determines if a pixel already has an entry in the FVC, if so, the FVC increases the corresponding frequency counter by one. However, because FVC size is limited, the FVC uses an eviction policy to determine which pixel frequencies to keep track of. Similar to a fully associative cache, the FVC uses the least frequent color (LRC) policy, where it evicts the pixel value with the smallest frequency when an entry is needed to track the frequency of a new pixel value.

\paragraph{\textbf{Hardware Cost}} Each FVC entry contains a color value (32 bits for RGBA), a validity flag (1 bit), and a counter with $log2$(number of screen pixels) bits. For example, a 64-entry FVC sized for a 4k$\times$4k display will only require 456 bytes of storage.

\subsection{The Common Colors Dictionary (CCD) }
\label{subsec:ccd}
CCD is used to encode compressed pixels. At the end of each frame, the FVC holds the frequencies of the estimated most common colors. The FVC is then used to construct the CCD for the next frame. Each CCD entry maps a pixel value to a dictionary (encoding) value. The CCD is implemented using a fully associative structure. 

When reading a surface, the mapping of CCD is reversed to decompress encoded pixels. We call the direct mapped structure that holds this reversed mapping the rCCD. Upon compressing a frame, or a set of frames, the rCCD mapping is attached to the frame and stored in main memory. Later on, when the frame is read, the rCCD is used to decompress the frame as described in Section~\ref{subsec:dcp_read} below.

\paragraph{\textbf{Hardware Cost}} CCD/rCCD with 64 entries only requires 264 bytes of storage.

\subsection{The Compression Status Buffer (CSB)}
Similar to other block-based compression algorithms~\cite{rasmusson2007exact,pool2012lossless,strom2008floating,hasselgren2006efficient}, a metadata buffer is used to hold the status of each compression block. For DCP, the CSB buffer indicates whether a given sub-block is compressed, where CSB holds one bit per sub-block. In our baseline, this translates to a cost of 1 bit per 128 bit of surface data. Both CSB and rCCD are needed to read a compressed frame as explained in the next section. 

\begin{figure}
	\centering
	\includegraphics[width=0.5\textwidth]{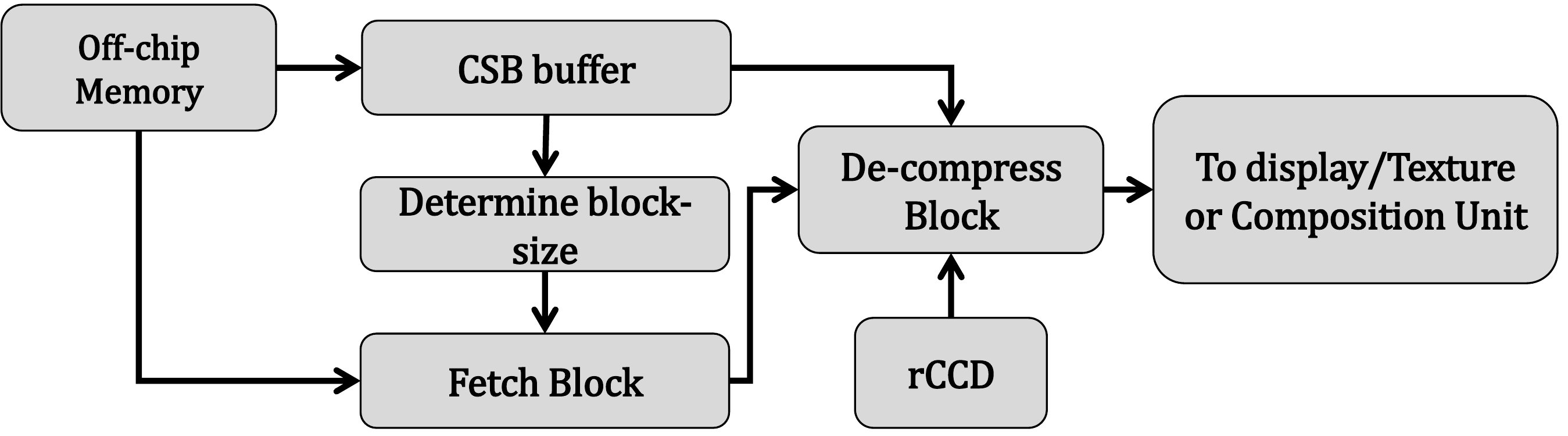}
	\caption{Reading a DCP compressed surface.}
	\label{fig:dcp_read}
\end{figure}

\subsection{Reading a Compressed Framebuffer Surface}
\label{subsec:dcp_read}
Figure~\ref{fig:dcp_read} shows the process of reading a compressed surface. It starts with loading the corresponding rCCD and CSB. To read a pixel, CSB entries are decoded to determine the size of compressed data and how many bytes should be fetched for each block. To avoid double latency, and since CSB size is relatively small, the CSB can be prefetched to a small on-chip buffer/cache. Once CSB is used to determine the size of a compressed block, the block is then fetched and the rCCD is used to decompress the values in each sub-block as shown in~\autoref{fig:dcp_read}.

\subsection{Multi-Surface Support}
\label{subsec:multisurface}

\textit{\textbf{Multiple Render Targets (MRT)}:} Some graphics applications may render to multiple target surfaces. Techniques that use MRT, like deferred shading, are popular in 3D applications and used to render scenes with complex lighting~\cite{unity3dDeferred}. To support multiple render targets, we need to replicate some of the structures in Figure~\ref{fig:dcp_stages} to match the maximum possible number of target surfaces. DCP will need a single FVC and a single CCD unit per render target. However, no need for additional FVC and CCD units if multiple passes are used to process MRT. 

Since most UI and 2D workloads render to a single target, a typical hardware implementation may only need support a single render target and the rare case of multiple targets is handled by using DCP with just a single surface. However, as discussed in Section~\ref{sec:hwcost}, adding extra structure is relatively cheap and cost little chip area.

\textit{\textbf{Multi-Surface Composition}:} Contemporary compositor engines can composite up to 16 surfaces in one pass~\cite{vivantecores}. To support multi-surface composition, the number of rCCD structures in Figure~\ref{fig:dcp_read} should match the number of surfaces that can be composited in parallel.

\subsection{Coupling DCP with Other Compression Algorithms}
\label{sec:multicomp}
DCP targets common UI and 2D applications. Other compression algorithms are better suited to 3D and some 2D applications. Industry practitioners have proposed supporting multiple compression algorithms~\cite{nvidiaxwhitepaper,kulshrestha2010selecting}. This means that in a hybrid scheme, each block can be compressed either using DCP or an alternative algorithm. In Section~\ref{subsec:hybrid} we evaluate the results of combining DCP with RAS.

\subsection{Dynamically Enabling DCP}
In this section, we explain how DCP can be enabled/disabled based on the expected compression performance. DCP performance can be predicted using the frequencies collected by the FVC at the end of a frame. By adding frequency values in the FVC, then comparing it to the total number of pixels (sample size), we can calculate what we call FVC coverage, which can be used to predict DCP performance, where:
\[
\text{FVC coverage} = \frac{\text{Sum of FVC frequencies}}{\text{Number of samples}}
\]

By defining a coverage threshold (CT) and comparing it to FVC coverage, then DCP can be used only if FVC coverage $\geq$ CT. By periodically enabling FVC, e.g., once every $n$ frames, FVC coverage can be updated and used to determine if DCP should be enabled. For an N-entry FVC, calculating coverage takes $N-1$ integer addition and one division operations per frame.
\begin{figure}
	\centering
	\includegraphics[width=0.45\textwidth]{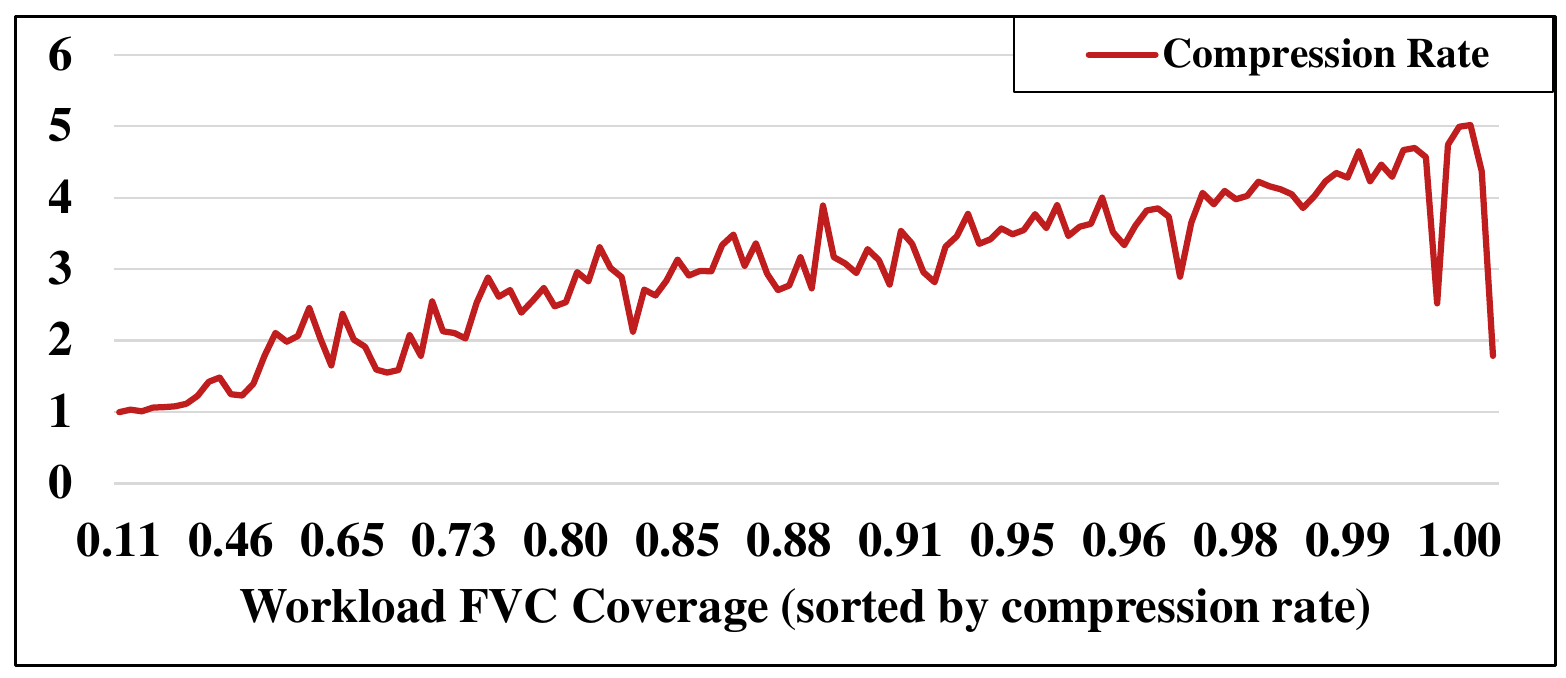}
	\caption{Compression rates vs. FVC coverage.}
	\label{fig:fvc_cov_comp}
\end{figure}

Figure~\ref{fig:fvc_cov_comp} shows FVC coverage vs. compression rates across workloads in Table~\ref{table:workloads}. It is clear that higher compression rates are achieved with higher FVC coverage. In our set of workloads, using DCP with FVC coverage $\geq$ 0.7 seems to achieve good compression rates ($>$ 2). 

Figure~\ref{fig:fvc_cov_comp} also shows some cases where larger FVC coverage yields lower compression. These cases represent workloads that exhibit sudden changes in frames, as a result, temporal coherence is lower than that of other benchmarks with similar FVC coverage. Two examples from Figure~\ref{fig:fvc_cov_comp} (the two large dips at the right end) are \emph{Unwind} which exhibits a UI with changing color brightness and \emph{Super Hexagon} which exhibits an interface that continuously switches theme colors.

\section{DCP Schemes}
\label{sec:dcp_schemes}
\subsection{Baseline DCP}
\label{subsec:basic_dcp}
In baseline DCP, the CCD is constructed using all FVC entries; thus, the number of entries in the CCD will always match FVC, and compressed blocks will have a fixed size of $log2$(FVC size)$\times$ (pixels per block) bits. 

\begin{figure}
	\centering
	\includegraphics[width=.33\textheight]{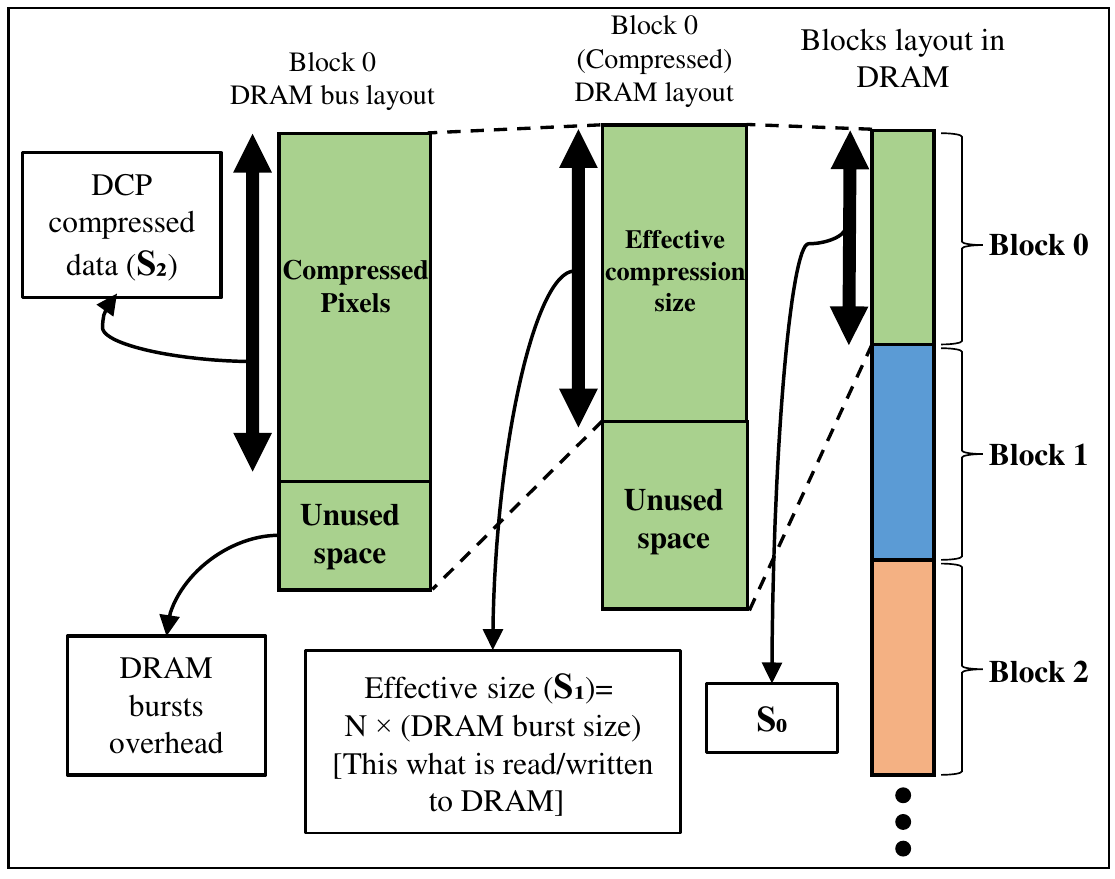}
	\caption{DCP memory Layout.}
	\label{fig:dramlayout}
\end{figure}

\paragraph{\textbf{Memory layout and effective compression rates}} Figure~\ref{fig:dramlayout} shows the memory layout of a DCP compressed surface. Space allocated to DCP blocks (0-2) is fixed (\textbf{$S_0$}, i.e., the size of an uncompressed block). On the other hand, the actual utilized space is determined by the size of compressed data (\textbf{$S_2$}). But because DRAM reads/writes data blocks using a number of bandwidth cycles that are burst size multiples, a block that should be compressed by $S_0/S_2$ will have an effective compression rate of $S_0/S_1$, where $S_1$ is the size of DRAM bursts needed to read compressed data.

In this work, we use the effective compression rate which reflects the reduction in memory bandwidth. In the remainder of this section, two variants of DCP (ADCP and VDCP) are introduced in addition to a hybrid scheme combining DCP and RAS (HDCP).

\subsection{Adaptive DCP (ADCP)}
\label{sec:adcp}
\begin{figure}
	\centering
	\includegraphics[width=.35\textheight]{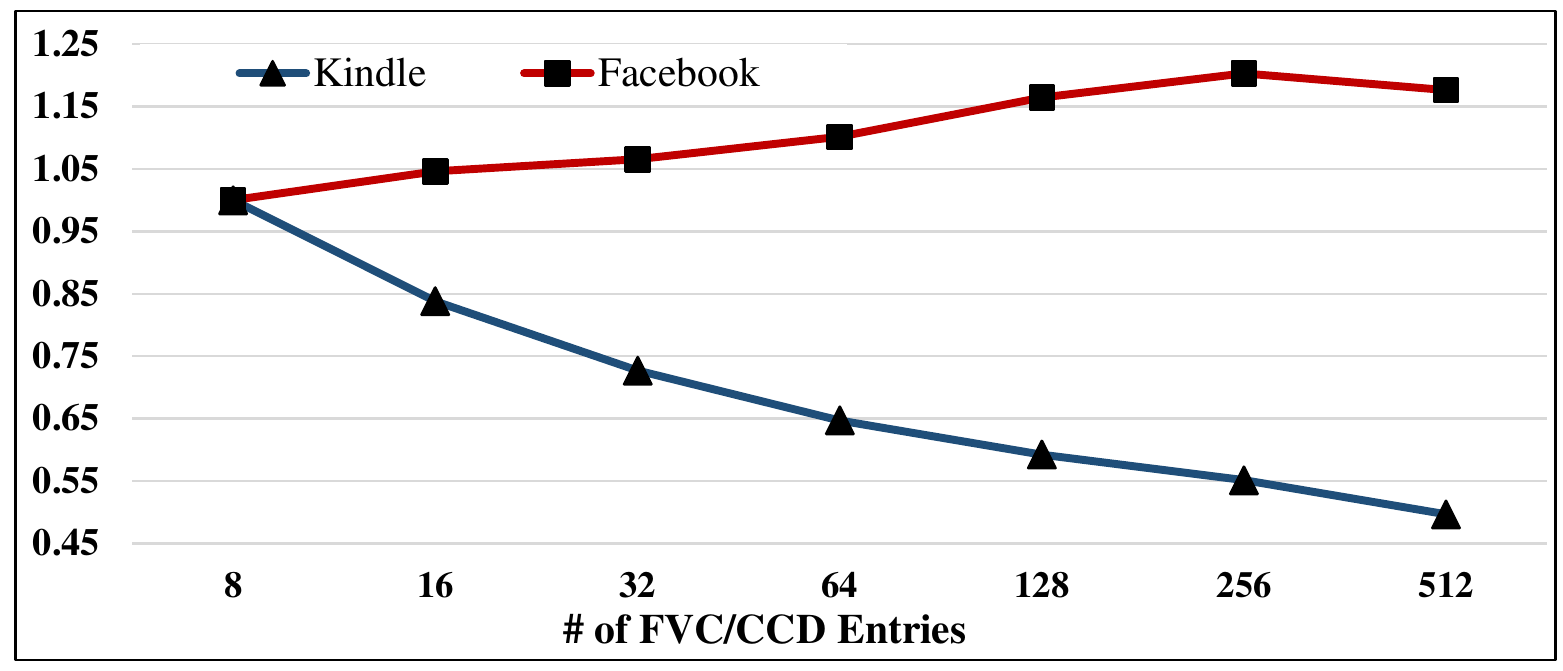}
	\caption{DCP compression vs. CCD size.}
	\label{fig:fvcSizeChange}
\end{figure}
ADCP is a variation of DCP that uses the distribution of frequent color values in the FVC to adjust the number of CCD entries. ADCP looks for the best trade-off between the number of compressible blocks and the size of their encoding. 

Frames of different applications, or different frames within the same application, may perform better/worse under larger/smaller palette sizes. Figure~\ref{fig:fvcSizeChange} shows DCP compression rates for \emph{Facebook} and \emph{Kindle} using 16 to 512 entry CCDs. \emph{Kindle} with its simple text achieves higher compression rates using smaller CCDs. On the other hand, \emph{Facebook} achieves the best compression rate using a 256-entry CCD. A larger CCD covers a wider range of values and it is able to compress more blocks, while a smaller CCD uses smaller encoding sizes.  For example, if a frame that uses 32-bit pixels with blocks that are 80\% white, 18\% blue, 1\% black and 1\% red uses a 2-entry CCD, 98\% of the blocks can be compressed using 1 bit per pixel for a total compression rate of 19.75$:$1 (ignoring metadata overhead). Another option is to use a 4-entry CCD to compress all the frame using 2 bits per pixel producing a compression rate of 16$:$1.

ADCP tries to optimize CCD size for each case by actively predicting the optimal number of CCD entries. CCD size determines encoding sizes and subsequently the size of compressed blocks.  ADCP uses FVC to predict the optimal CCD size using Algorithm~\ref{algo:adcp}. In Algorithm~\ref{algo:adcp}, FVC frequencies, sorted from most to least frequent in FVC\_Val, are used as input. Note that to simplify calculations, DRAM burst size and pixels layout were ignored.  

ADCP has a negligible overhead; the number of iterations in Algorithm~\ref{algo:adcp} depends on the number of FVC entries. For example, for 64-entry FVC, the loop will only execute six times (i.e., $log2$(FVC size)).

\algnewcommand{\LineComment}[1]{\State \(\triangleright\) #1}
\begin{algorithm}
	\caption{Predicting optimal CCD size}
	\label{algo:adcp}
	\begin{algorithmic}
		\small
		\State \textbf{INPUTS} (FrameSizePixels, PixelSizeBits, FVC\_Val, Max\_FVC\_Size)
		\LineComment{predicted compressed frame size in bits}
		\State expected\_frame\_size = Frame\_W$\times$H*PixelSizeBits
		\LineComment{Optimal CCD entries = 2$^{\text{opt\_CCD}}$}
		\State opt\_CCD = 0 
		\For {$i$=0 to $log2$(Max\_FVC\_Size)}
		\State sum = SumFrequencies(FVC\_Val(0) to FVC\_Val($2^i$-1)) 
		\State frame\_size = sum * i + (FrameSizePixels-sum)* PixelSizeBits
		\State \If {frame\_size $<$ expected\_frame\_size}
		\State expected\_frame\_size = frame\_size
		\State opt\_CCD = i    
		\EndIf
		\EndFor
		\State \textbf{return} 2$^{\text{opt\_CCD}}$
	\end{algorithmic}
\end{algorithm}

\subsection{Variable DCP (VDCP)}
\label{sec:vdcp}
\begin{figure}
	\centering
	\includegraphics[width=.4\textwidth]{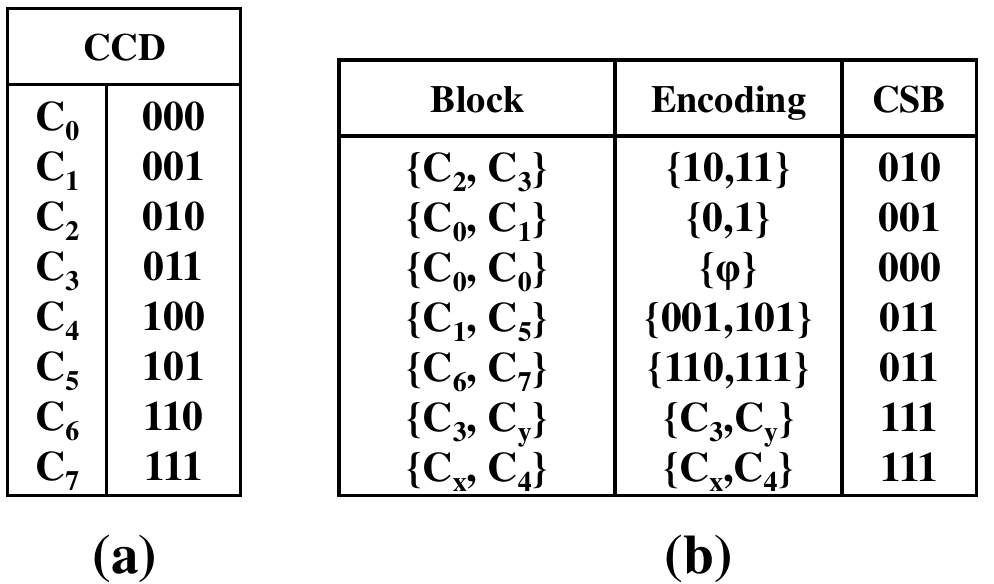}
	\caption{VDCP example encoding (a). CCD entries (b). The ecoding of 2-pixel blocks using the CCD in Table~(a).}
	\label{fig:adcpexample}
\end{figure}

VDCP is another DCP variation that goes further than ADCP by adapting palette sizes to optmize compression at the sub-block level. VDCP uses variable-length coding by changing the number of rCCD entries used to encode/decode each sub-block. VDCP reduces the number of encoding bits per pixel to $i = ceil( log_2( max (pixel\_color\_index)))$, which means that $i$ is determined by the pixel within the sub-block that has the highest index (i.e., lowest frequency) in the CCD. 

With VDCP, CSB is used to determine the number of rCCD entries used for each sub-block, where the number rCCD entries equals to $2^{CSB\_Value}$ (i.e., encoded colors fall in the first $2^{CSB\_Value}$ CCD entries), and a special CSB\_Value is used for uncompressed sub-blocks. 

Figure~\ref{fig:adcpexample} shows a VDCP example. In Figure~\ref{fig:adcpexample}.a, an example CCD is shown, where the most frequent color value, $C_0$, is encoded to 000 and the least frequent value, $C_7$, is encoded to 111. Figure \ref{fig:adcpexample}.b shows the VDCP encoding for seven 2-pixel sub-blocks. As shown in the figure, the CSB tracks each sub-block's encoding. 000$_2$ in the CSB indicates that only the most frequent color in the CCD $C_0$ is used, while 001$_2$ indicates that the top 2 CCD colors, \{$C_0$ and $C_1$\}, are used and so on. 111$_2$ is used for uncompressed sub-blocks. 

In Figure~\ref{fig:adcpexample}.b, the first row shows a sub-block with values C$_2$ and C$_3$, this means that only the top $2^2$ entries in the CCD are used for encoding the sub-block. Subsequently, the corresponding CSB entry is set to 010$_2$, and each pixel color is encoded using 2 bits. The second sub-block in Figure\ref{fig:adcpexample}.b contains \{C$_0$,C$_1$\}, encoded using the top $2^1$ CCD entries (1 bit per color), and the corresponding CSB entry is set to 001$_2$. The third sub-block contains only $C_0$, and the corresponding CSB value in this case is 000$_2$, which indicates the content for the entire sub-block (since only the top entry in the table is used). Note that to make this example easy to follow, the CCD is shown with eight entries (instead of 64), so the CSB values 100$_2$ to 110$_2$ are not in use.

\subsection{Hybrid DCP (HDCP)}
\label{subsec:HDCP}
DCP is only effective on a subset of applications. Ideally, it should be used with other compression algorithms. We evaluate a Hybrid DCP that combines DCP with RAS. HDCP compresses each block using DCP and RAS and uses the result with higher compression rate. To support the additional compression modes, the number of CSB bits is increased. Results in Section~\ref{sec:methres} show that this technique produces higher compression rates at the cost of additional on-chip computations.

\subsection{DCP Implementation}
\label{subsec:implementation}

In addition to hardware structures (FVC, CCD and rCCD), DCP requires some support from the software layer. To implement DCP, the graphics driver will attach DCP data as part of the state associated with a surface (along other state data like size and formatting). For VDCP, Algorithm~\ref{algo:adcp} can be added to the driver as well, where it can calculate next CCD size at the end of each frame.

\section{Methodology} 
\label{sec:methres} 

\begin{table}[t] 
	\small 
	\center 
	\begin{tabular}{|l|c|} 
		\hline 
		\multicolumn{2}{|c|}{ \textbf{Baseline DCP Configurations}}\\\hline 
		FVC Size & 64 entries \\ \hline 
		FVC replacement policy & Least-frequent value \\ \hline 
		CCD size & 64 entries \\ \hline 
		Pixel Block size & 8$\times$ 8 \\ \hline 
		Pixel Sub-block size & 2$\times$2 \\ \hline 
		\multirow{2}{*}{CSB bits per sub-block} & 1 (DCP, ADCP \& HuffDCP) \\ 
		& 3 (VDCP), 5 (HDCP) \\ \hline 
		Memory Burst Size & 128 bits \\ \hline 
		Pixel Sampling Rate & 1:1 \\ \hline 
	\end{tabular} 
	\caption{Baseline Configurations} 
	\label{table:fconfigs} 
\end{table}

\begin{table*}[t] 
	\scriptsize	 
	\center 
	\resizebox{\textwidth}{!}{
	\begin{tabular}{|l|l|l||l|l|l||l|l|l||l|l|l||l|l|l|} 
		\hline 
		\# & Cat. & Benchmark & \# & Cat. & Benchmark & \# & Cat. & Benchmark & \# & Cat. & Benchmark & \# & Cat. & Benchmark\\ \hline \hline 
		
		1    &     UI     &     Android Settings     &    26    &     UI     &     Pocket     &    51    &     UI     &     Yellowpages     &    76    &     UI     &     Textra     &    101    &     2D     &     Unwind     \\ \hline
		2    &     UI     &     Morecast     &    27    &     UI     &     ES File Explorer     &    52    &     UI     &     Eye in the Sky     &    77    &     UI     &     WPS Office     &    102    &     2D     &     Color Switch     \\ \hline
		3    &     UI     &     Poweramp     &    28    &     UI     &     Chrome     &    53    &     UI     &     OfficeSuite     &    78    &     UI     &     People Contacts     &    103    &     2D     &     Impossible Game     \\ \hline
		4    &     UI     &     Speedest     &    29    &     UI     &     Applock     &    54    &     UI     &     Dictionary.com     &    79    &     UI     &     Unit Conv. Ult.     &    104    &     2D     &     Flow     \\ \hline
		5    &     UI     &     Twitter     &    30    &     UI     &     Accuweather     &    55    &     UI     &     Walgreens     &    80    &     UI     &     Skyscanner     &    105    &     2D     &    2048    \\ \hline
		6    &     UI     &     Facebook     &    31    &     UI     &     Flipboard     &    56    &     UI     &     Walmart     &    81    &     UI     &     Calendar     &    106    &     2D     &     Gyro     \\ \hline
		7    &     UI     &     Twitch     &    32    &     UI     &     Booking.com     &    57    &     UI     &     CNN     &    82    &     UI     &     Merriam Webster     &    107    &     2D     &     99 Problems     \\ \hline
		8    &     UI     &     Wish     &    33    &     UI     &     Shazam     &    58    &     UI     &     File Commander     &    83    &     UI     &     ESPN     &    108    &     2D     &     Dumb Ways to Die     \\ \hline
		9    &     UI     &     Imgur     &    34    &     UI     &     Zedge     &    59    &     UI     &     Terminal Emulator     &    84    &     UI     &     Tumblr     &    109    &     2D     &     Piano Tiles     \\ \hline
		10    &     UI     &     Soundcloud     &    35    &     UI     &     Indeed     &    60    &     UI     &     Adobe Acrobat     &    85    &     UI     &     Quickpic     &    110    &     2D     &     loop     \\ \hline
		11    &     UI     &     Automate     &    36    &     UI     &     Runkeeper     &    61    &     UI     &     Android Call     &    86    &     UI     &     Duolingo     &    111    &     2D     &     Ultraflow     \\ \hline
		12    &     UI     &     Musixmatch     &    37    &     UI     &     Steam     &    62    &     UI     &     Gallery     &    87    &     UI     &     Clock     &    112    &     2D     &     Okay     \\ \hline
		13    &     UI     &     Airbnb     &    38    &     UI     &     Khan Academy     &    63    &     UI     &     Feedly     &    88    &     UI     &     Google Messenger     &    113    &     3D     &     Traffic Rider     \\ \hline
		14    &     UI     &     CBS Sports     &    39    &     UI     &     The Weather Channel     &    64    &     UI     &     Baconreader     &    89    &     UI     &     Calculator     &    114    &     3D     &     Extreme Car Driving     \\ \hline
		15    &     UI     &     Etsy     &    40    &     UI     &     Yahoo Finance     &    65    &     UI     &     aCalendar     &    90    &     UI     &     Soundhound     &    115    &     3D     &     3D Bowling     \\ \hline
		16    &     UI     &     Android Home     &    41    &     UI     &     Tapatalk     &    66    &     UI     &     Bakareader     &    91    &     UI     &     Translate     &    116    &     3D     &     Dr. Driving     \\ \hline
		17    &     UI     &     Pinterest     &    42    &     UI     &     Kickstarter     &    67    &     UI     &     Kindle     &    92    &     UI     &     Any.do     &    117    &     3D     &     Paper Toss     \\ \hline
		18    &     UI     &     Aldiko     &    43    &     UI     &     Amazon Store     &    68    &     UI     &     eBay     &    93    &     2D     &     Candy Crush Saga     &    118    &     3D     &     Rolling Sky     \\ \hline
		19    &     UI     &     Letgo     &    44    &     UI     &     Zomato     &    69    &     UI     &     Venmo     &    94    &     2D     &     Trainyard     &    119    &     3D     &     Stack     \\ \hline
		20    &     UI     &     Yelp     &    45    &     UI     &     Spotify     &    70    &     UI     &     Mcdonalds     &    95    &     2D     &     Mines     &    120    &     3D     &     Zigzag     \\ \hline
		21    &     UI     &     Android Messaging     &    46    &     UI     &     Runtastic     &    71    &     UI     &     Colornote     &    96    &     2D     &     Cut the Rope 2     &    121    &     3D     &     Stargather     \\ \hline
		22    &     UI     &     BBC iPlayer     &    47    &     UI     &     theScore     &    72    &     UI     &     Reddit     &    97    &     2D     &     Angry Birds     &    122    &     3D     &     Commute H. Traffic     \\ \hline
		23    &     UI     &     Tachiyomi     &    48    &     UI     &     Food Network     &    73    &     UI     &     Checkout 51     &    98    &     2D     &     Strata     &    123    &     3D     &     Crossy Road     \\ \hline
		24    &     UI     &     gReader     &    49    &     UI     &     MX Player     &    74    &     UI     &     Tasker     &    99    &     2D     &     Brain it On     &    124    &     3D     &     Smashy Road     \\ \hline
		25    &     UI     &     Google Maps     &    50    &     UI     &     VLC     &    75    &     UI     &     IFTTT     &    100    &     2D     &     Super Hexagon     &        &        &        \\ \hline

	\end{tabular}}
	\caption{List of Android workloads} 
	\label{table:workloads2} 
\end{table*}

\begin{table}[t] 
	\small 
	\center 
	\begin{tabular}{|l|c|p{35mm}|} 
		\hline 
		\multicolumn{3}{|c|}{\textbf{System Configurations}}\\ \hline 
		\multicolumn{2}{|c|}{Operating System} & Android 4.2.2 (API 17)\\ \hline 
		\multicolumn{2}{|c|}{Display Size}& 720$\times$1280 \\ \hline \hline 
		\multicolumn{3}{|c|}{\textbf{Android workloads}} \\ \hline 
		\multicolumn{2}{|c|}{\textbf{Category}} & \textbf{\# of workloads} \\ \hline 
		\multicolumn{2}{|c|} {UI Applications} & 92 (24031 frames) \\ \hline 
		\multicolumn{2}{|c|} {2D Applications }& 20 (7888 frames) \\ \hline 
		\multicolumn{2}{|c|} {3D Applications }& 12 (2549 frames) \\ \hline 
		\multicolumn{2}{|c|} {\textbf{Total \# of Applications }}& \textbf{124 (34468 frames)} \\ \hline 
	\end{tabular} 
	\caption{System configurations and workloads summary} 
	\label{table:workloads} 
\end{table} 

Our experimentation configurations are listed in Table~\ref{table:fconfigs}. We calculated compression rates using a model that assumes a tile-based GPU architecture. Our model works as follows: 
\begin{itemize}
	\item First, we feed the frames of each workload to our model, which then splits each frame to 8x8 blocks.
	
	\item For each block, the model calculates the compressed size of each sub-block. The total of compressed and uncompressed sub-block sizes are added to calculate the compressed size of the block.
	
	 \item Compressed block size is then used to calculate the number of DRAM bursts required. The model then calculates the total bandwidth consumed by a compressed frame by summing the number of DRAM bursts of all the blocks in the frame.
\end{itemize}

Note that the model computes compression rates starting from the second frame, using the first frame to populate the first FVC and CCD.

We evaluated surface compression using our set of randomly chosen popular Android applications (Table~\ref{table:workloads2}). Our traces will be published and made available for any future studies. 

We split applications into three groups: UI applications, 2D applications, and 3D games. All of our benchmarks use OpenGL ES and render to a single target buffer (up to OpenGL ES 2.0 MRT is only supported through vendor extensions~\cite{oglesMrtExt}). 

We manually interacted with each application to execute a simple task. In total, we used 34468 frames that represent 124 applications (shown in Table~\ref{table:workloads}). We only consider regions of interest in each workload that represent the typical use case of the workload (i.e., loading/initialization frames are not considered).

The rest of configurations are listed in Table~\ref{table:fconfigs}. The effective compression rate and metadata overhead are taken into account when calculating the total compression rate. We use a block size of 8$\times$8 pixels (256 bytes), which matches the block sizes used by RAS.


In addition to DCP, we evaluate two lossless methods described in Section~\ref{sec:relatedWork}. \emph{RED} uses Nvidia's compression \cite{nvidiaxwhitepaper} and \emph{RAS}, which is based on work of Rasmusson et al. \cite{rasmusson2007exact}.
\emph{RAS} is a prediction based algorithm that predicts the value of a pixel using neighbor pixel values. The difference between prediction and the actual value is then encoded using Golomb-Rice coding. We used parameters suggested by Rasmusson et al. \cite{rasmusson2007exact}, namely 8$\times$8 blocks and, as described in the paper, we set the value of the Golomb-Rice parameter k by exploring values between 0 and 6, use k = 7 for the ``special mode'', and use the suggested ``3 sizes mode'' for higher compression rates. We organize color values by their color channel as described in Str{\"o}m et al.~\cite{strom2008floating}. We experimented with RAS using RGBA and $Yc_oc_g$ formats and found that for many applications, particularly UI and 2D, RAS shows favorable results using RGBA channels. So we used RAS with RGBA channels in our comparison. 

For CSB, DCP and ADCP use 1-bit per sub-block. VDCP uses 3 bits per sub-block; with an FVC size of 64, seven combinations are used--1, 2, 4, 8, 16, 32 and 64, plus a combination for non-compressed sub-blocks. To compare against techniques that use Huffman coding~\cite{shim2005frame}, a Huffman coded DCP (HuffDCP) is implemented, where FVC frequencies are used to construct CCD with variable length Huffman coding.

\section{Results and Discussion} 
\label{subsec:results}

\begin{figure*}[t!p] 
	\begin{subfigure}[b]{1\textwidth} 
		\centering 
		\includegraphics[width=.85\textwidth]{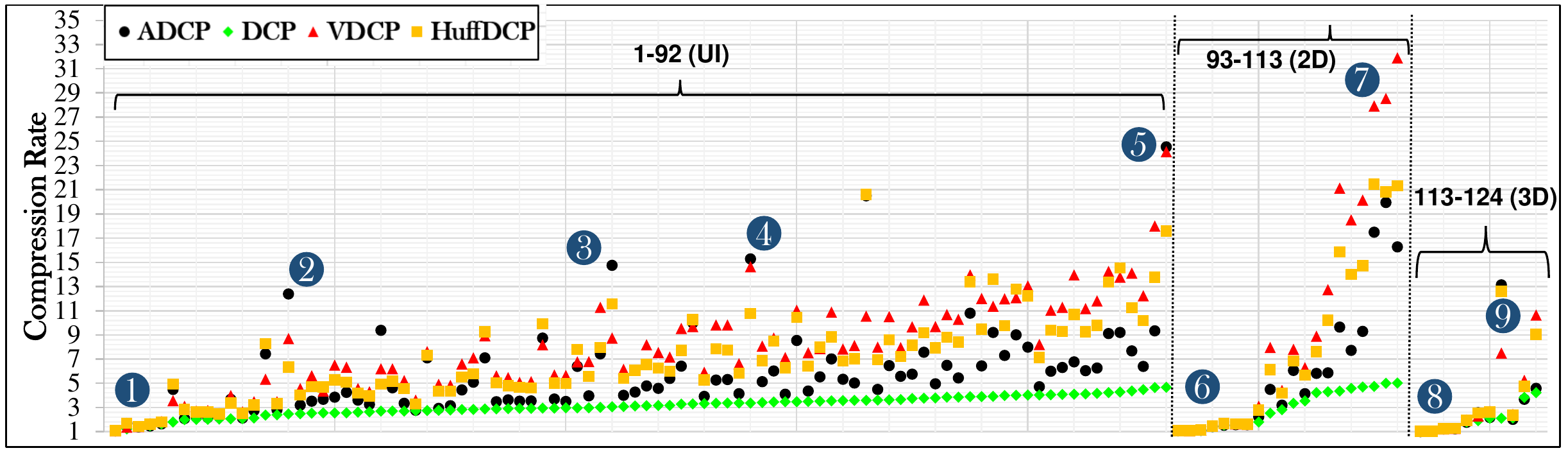} 
		\caption{DCP schemes compression.} 
		\label{fig:dcp_results1} 
	\end{subfigure} 
	
	\begin{subfigure}[b]{1\textwidth} 
		\centering 
		\includegraphics[width=.85\textwidth]{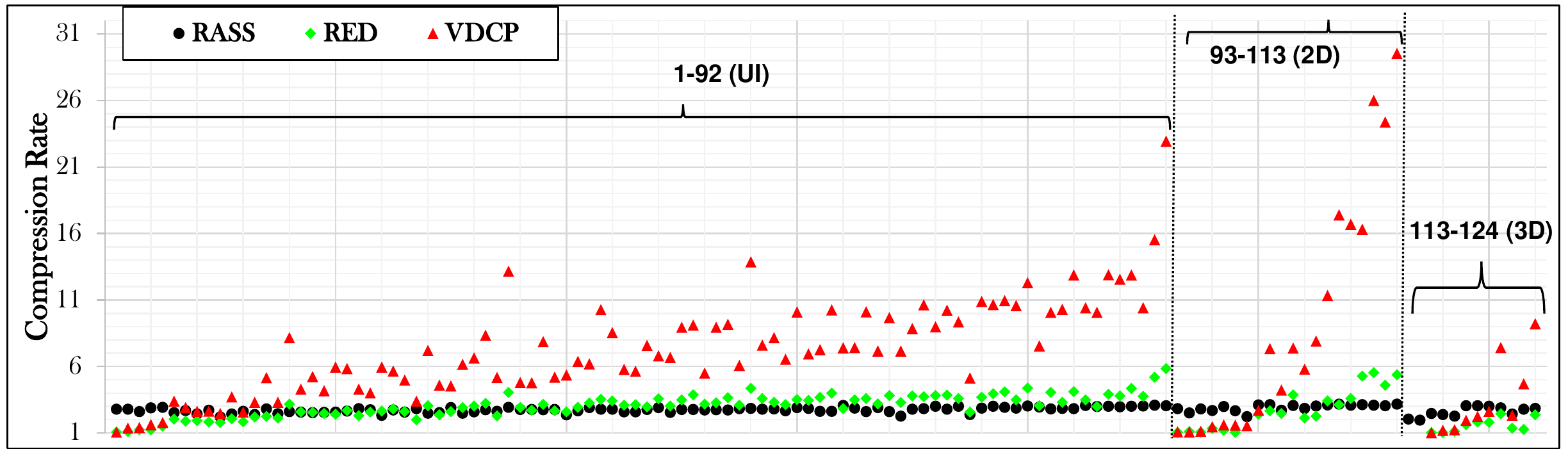} 
		\caption{Comparing RAS, RED, and VDCP compression rates.} 
		\label{fig:dcp_results2} 
	\end{subfigure} 
	\caption{Compression rates of workloads ordered from left to right following their order in Table~\ref{table:workloads2}.} 
	\label{fig:dcp_results} 
\end{figure*}

\begin{figure} 
	\centering 
	\includegraphics[width=0.42\textwidth]{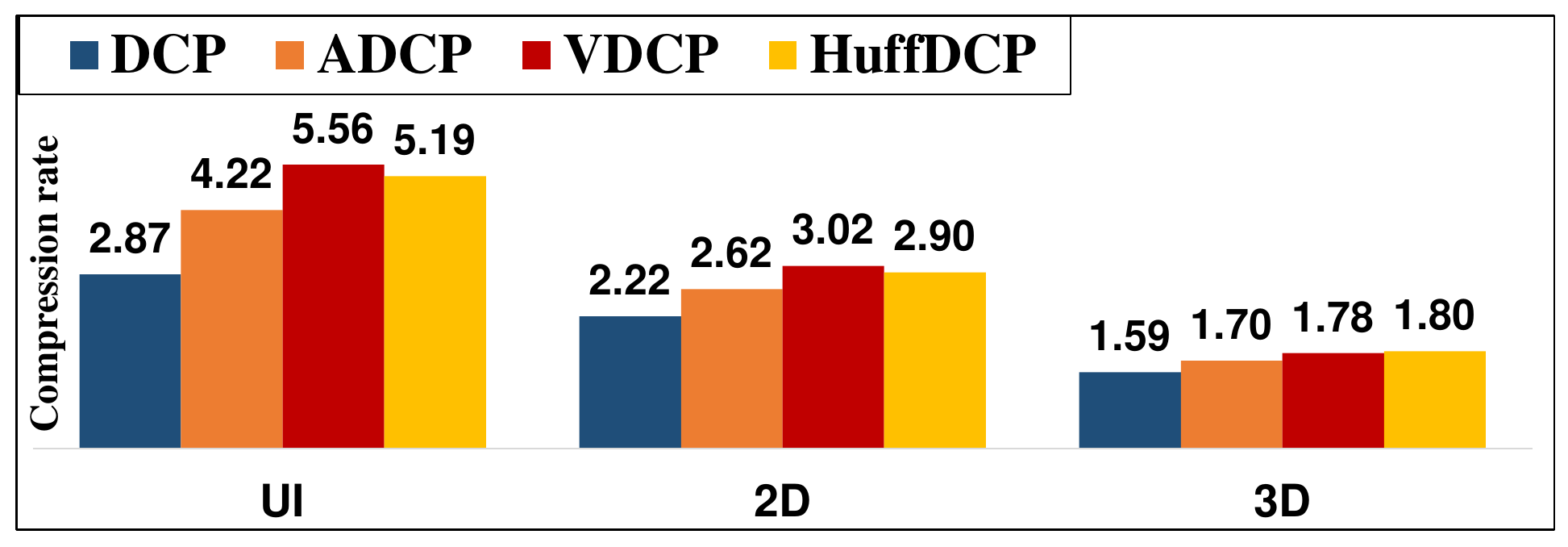} 
	\caption{Harmonic mean of DCP schemes compression rates per application category.} 
	\label{fig:dcp_results_summary} 
\end{figure} 

\begin{figure} 
	\centering 
	\includegraphics[width=0.42\textwidth]{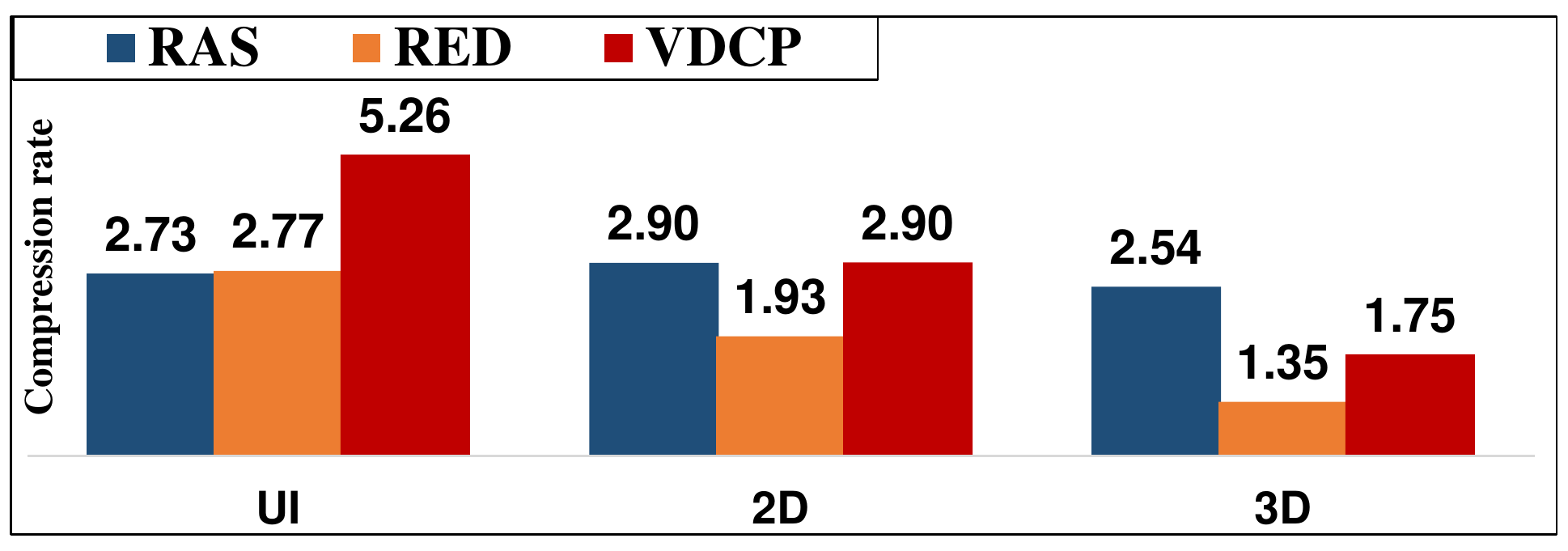} 
	\caption{Comparing RAS, RED and VDCP effective compression rates.} 
	\label{fig:dcp_others} 
\end{figure} 

\subsection{DCP Schemes} 

To compare DCP schemes, we isolate the effect of memory burst size and only take into account CSB overhead. Later, bursts are taken into account when comparing DCP, RAS and RED. We compare baseline DCP  against ADCP, VDCP and HuffDCP. 

Figure~\ref{fig:dcp_results1} shows compression rates in each category sorted by baseline DCP compression rate (the same order used in Table~\ref{table:workloads2}). The figure shows that the baseline DCP is the least effective scheme. UI applications~\circledB{1}, such as Android Settings, Morecast, and Poweramp, show low compression rates of less than 2. After examining these applications, we found that they feature gradient backgrounds and graphical elements that DCP cannot compress using small palettes. 

In~\circledB{2} (Zedge) and ~\circledB{3} (Spotify), ADCP achieves higher compression rates than HuffDCP and VDCP. Looking at these applications we found that they contain a mix of solid backgrounds and frames that contain images which DCP will, mostly, not be able to compress. ADCP can compress frames with solid backgrounds with lower overhead than VDCP since it has a lower CSB overhead. On the other hand, in frames containing images, both ADCP and VDCP will not be able to perform well, but ADCP will incur lower CSB overhead. 

For applications with simple color schemes, such as OfficeSuite \circledB{4} and Any.do \circledB{5}, VDCP and DCP achieve high compression rates. Nevertheless, VDCP, ADCP, and HuffDCP were all able to achieve even higher compression rates. Looking at 2D applications, performance varies significantly. 

In \circledB{6}, applications with sophisticated graphics, like Candy Crush, Trainyard, Mines, Cut the Rope and Angry Birds, have low compression rates ($<$ 1.7). On the other hand, applications using simpler graphics (e.g., loop Ultraflow, and Okay) achieve high compression rates, especially with VDCP \circledB{7}. A similar trend is exhibited in \circledB{8}, where graphically rich 3D games (Traffic Rider, Extreme Car Driving, and 3D Bowling) show low compression rates. On the other hand, games like Smashy Road, show good compression rates (highest VDCP at 10.63). Also Stargather \circledB{9}, with similar characteristics to UI applications in \circledB{2} and \circledB{3}, shows higher rates with HuffDCP and VDCP. 

Figure~\ref{fig:dcp_results_summary} summarizes the results in Figure~\ref{fig:dcp_results1}. VDCP shows better compression rates for UI and 2D applications with 5.56 and 3.02 respectively. For 3D games, HuffDCP shows the highest rate (1.80). HuffDCP and VDCP do better with 3D workloads since their compression rates are similar to VDCP but with lower CSB overhead. Interestingly, using Huffman encoding in HuffDCP achieves lower compression rates than VDCP in UI and 2D workloads. This is due to Huffman inefficiencies with probability distributions that are not exact powers of two. For example, if we have 32 bit values and frequencies of A(49.5\%), B(49.5\%), C(0.5\%) and D(0.5\%) then Huffman encoding will assign codes of 1 bit to A, 2 bits to B and 3 bits to C and D with a total compression rate of 21.12. ADCP and VDCP encode A and B using 1 bit, while keeping C and D uncompressed, resulting in a compression rate of 24.4. 



\subsection{Comparing VDCP, RAS and RED} 

Figure~\ref{fig:dcp_results2} compares VDCP with RAS and RED and Figure~\ref{fig:dcp_others} summarizes the results in Figure~\ref{fig:dcp_results2}. Memory bursts and CSB overhead were taken into account. For UI applications, VDCP achieves a mean effective compression rate of 5.26 compared to 2.73 for RAS and 2.77 for RED. VDCP performs well with UI and 2D applications. On the other hand, RAS, a more generic compression algorithm, has consistent performance across all workloads. RAS outperforms VDCP in 3D games (2.54, compared to 1.75 for VDCP).
Similar to VDCP, RED performs well with UI and 2D applications, but with lower rates that VDCP. VDCP performance with 3D workloads is the reasoning behind suggesting a hybrid approach consisting of DCP and another general purpose compression algorithms--similar to what is described some implementations~\cite{nvidiaxwhitepaper,kulshrestha2010selecting}. A Hybrid VDCP-RAS scheme is discussed in Section~\ref{subsec:hybrid}.

\subsection{\normalsize{Factors affecting FVC Fidelity}} 
\label{subsec:fvcexpr} 

In this section we discuss and quantitatively evaluate four factors that affect FVC and should be considered when using DCP.  

\paragraph{\textbf{FVC Size:}} Larger FVC sizes can capture frequent colors more accurately, as they are less likely to evict a frequent value from the FVC because of capacity. 

To evaluate how FVC size affect accuracy, we use \emph{relative coverage}. For an $N$-entry FVC, we calculate relative coverage by dividing the number of pixels represented by the $N$ top colors collected by FVC by the number of pixels represented by the actual $N$ most frequent colors. 

\begin{figure}[t!p] 
	\centering
	\begin{subfigure}[b]{0.40\textwidth} 
		\centering 
		\includegraphics[width=1\textwidth]{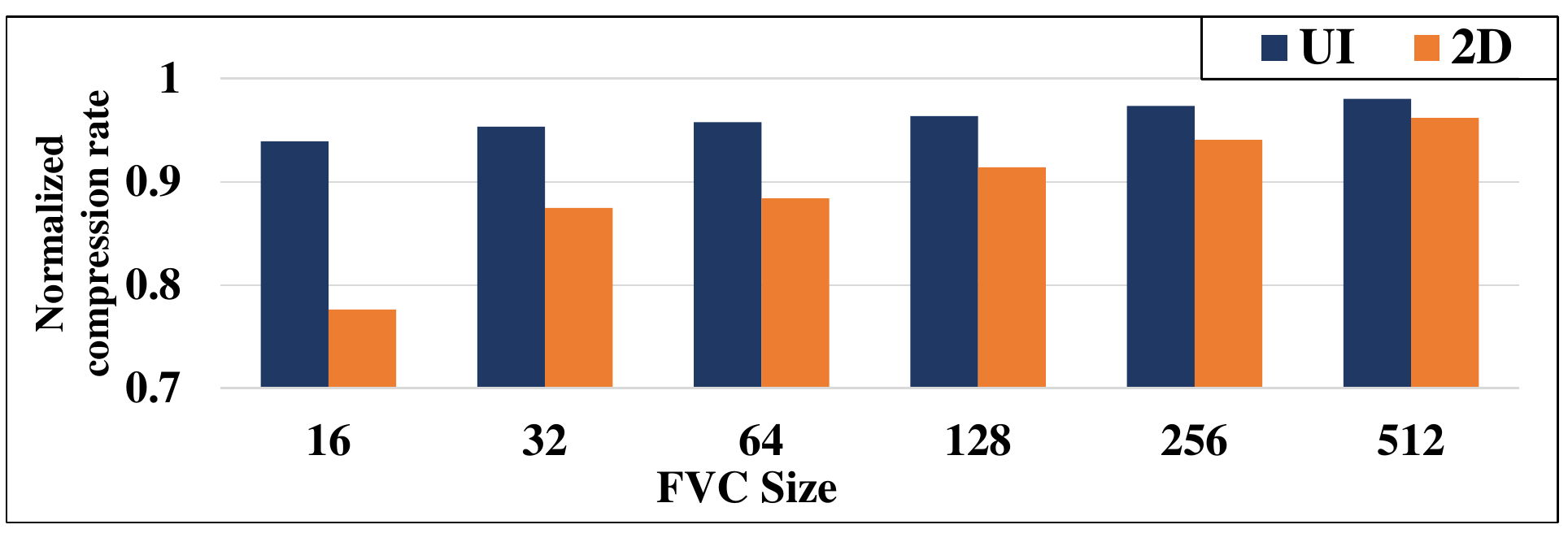} 
		\caption{FVC size vs. Relative Coverage.} 
		\label{fig:fvcres} 
	\end{subfigure} 
	
	\begin{subfigure}[b]{0.40\textwidth} 
		\centering 
		\includegraphics[width=1\textwidth]{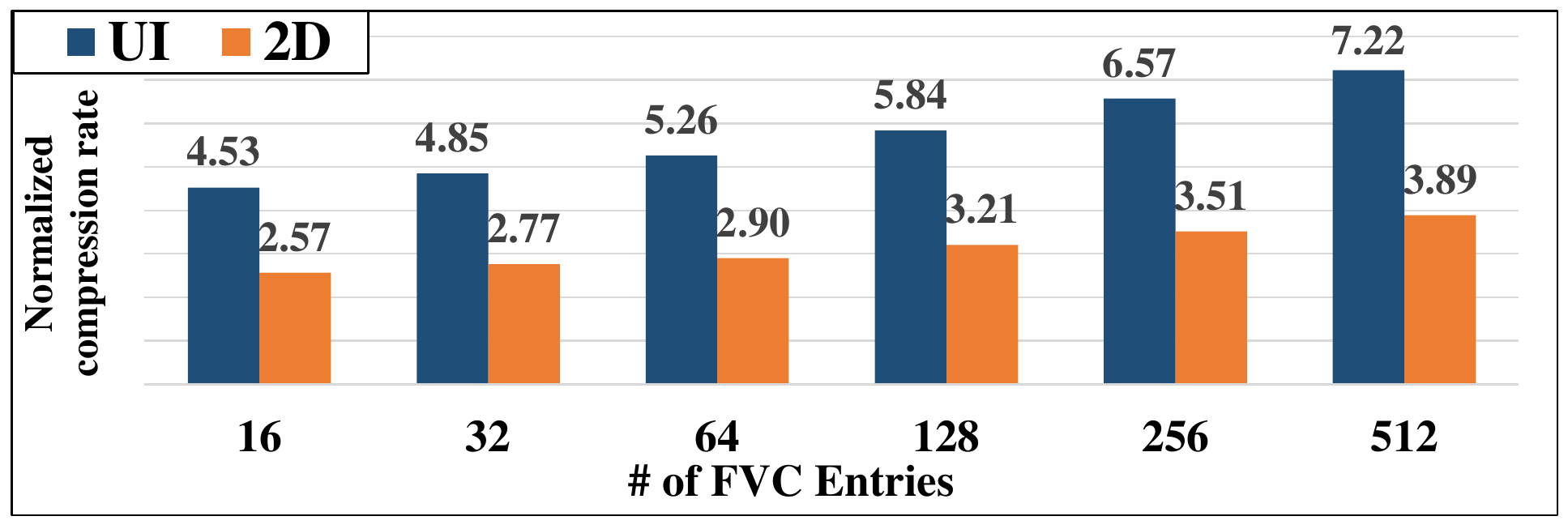} 
		\caption{FVC size vs. VDCP compression rate.} 
		\label{fig:fvc_compr} 
	\end{subfigure} 
	\caption{Comparing FVC size with relative coverage (a) and its effect on compression rates (b).} 
	\label{fig:fvc_size} 
\end{figure}

Figure \ref{fig:fvcres} shows the effect of FVC size for UI applications. A 16-entry FVC has a relative coverage of 94\% compared to 98.3\% for 512-entry FVC. This means a 16-entry FVC is able to capture colors that cover 94\% of the area covered by the actual 16 most frequent colors, while the 512-entry FVC is able to capture 98\% of the coverage the actual 512 most common colors are able to cover. Figure~\ref{fig:fvc_compr} shows how accuracy affect compression rates, as frequencies collected using larger FVCs are a better representation of the actual most common colors.

\begin{figure} 
	\centering 
	\includegraphics[width=.40\textwidth]{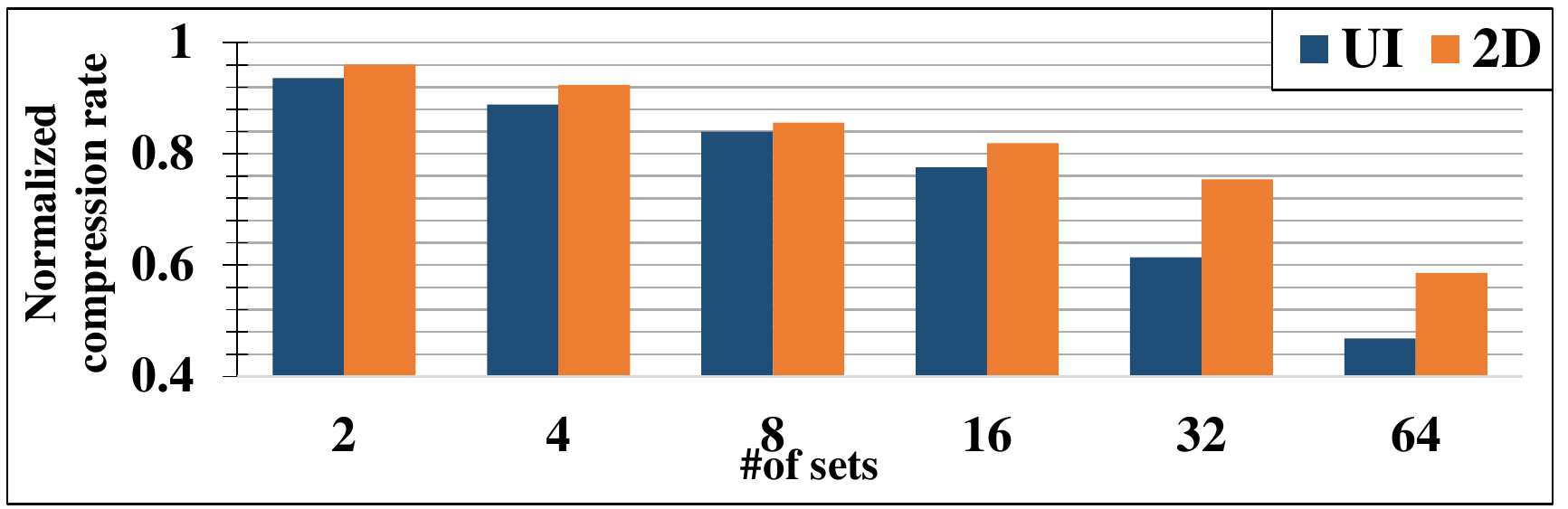} 
	\caption{64-entry FVC associativity vs. the compression rate of fully associative FVC.} 
	\label{fig:fvc_assc} 
\end{figure} 

\paragraph{\textbf{Replacement Policy and Associativity}}
We evaluated using a number of replacement policies: the baseline least-frequent color (LFC), second least-frequent (2LFC), least-recently-used (LRU), and random replacement. The idea behind including a 2LFC is to see the effect of avoiding thrashing newly discovered colors that are prone to eviction. 

 Using UI workloads with 64-entry FVC, the mean compression rate with LFC is 5.26, while it is 5.25 for 2LFC. On the other hand, LRU and random achieve lower rates of 3.04 and 2.92, respectively. 
 
 We also evaluated changing FVC associativity from fully associative to direct-mapped, and used color channel values to determine the set. As expected, the FVC performance degrades as we increase the number of sets (as shown Figure~\ref{fig:fvc_assc}).

\paragraph{\textbf{Pixel Sampling}}
We noticed that the FVC can be constructed using a subset of frame pixels, i.e., by sampling them using only one in every $n$th pixel to collect frequent colors statistics.

\begin{figure} 
	\centering 
	\includegraphics[width=.42\textwidth]{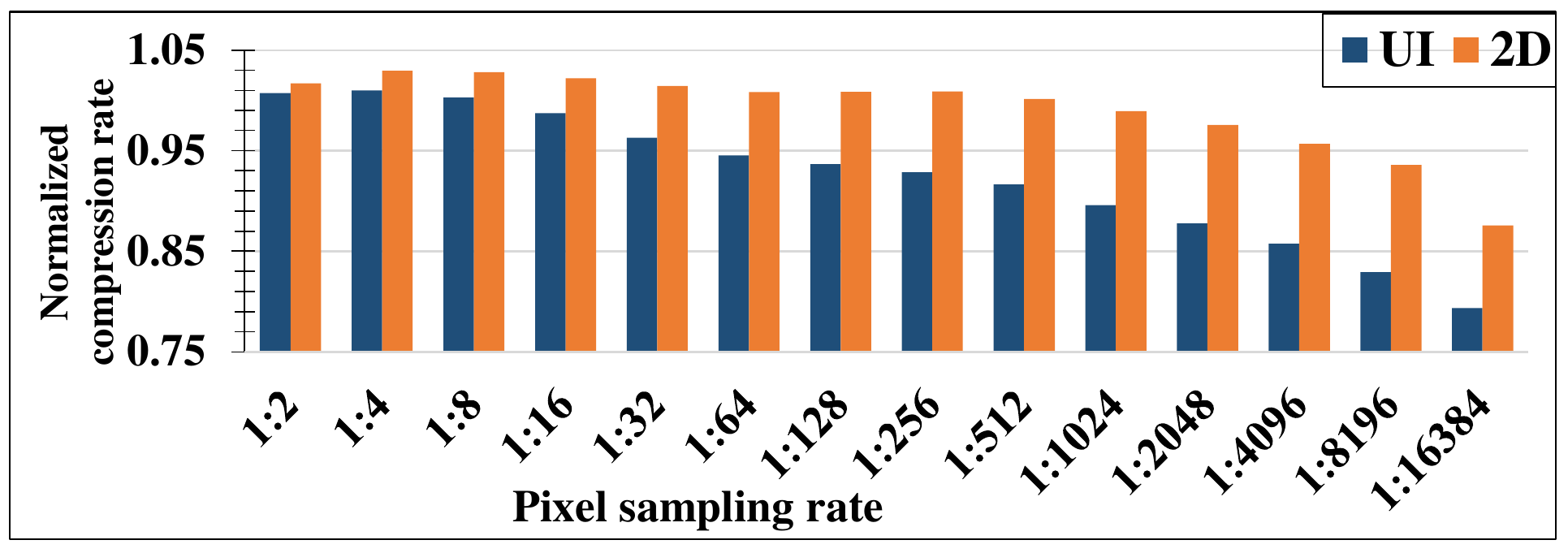} 
	\caption{Normalized compression rates vs. FVC pixel sampling rate for UI applications.} 
	\label{fig:pixelsampling} 
\end{figure}

Figure~\ref{fig:pixelsampling} illustrates the effect of pixel sampling on VDCP. We evaluate sampling rates from 1:1 (every pixel accesses the FVC) to 1:16384. 1:16 sampling achieves 98.7\% (UI) and 102\% (2D) of the compression achieved by 1:1 sampling. We expect that the slightly higher compression rate for 2D workloads is caused by sampling working as a noise filter.

\begin{figure} 
	\centering 
	\includegraphics[width=.42\textwidth]{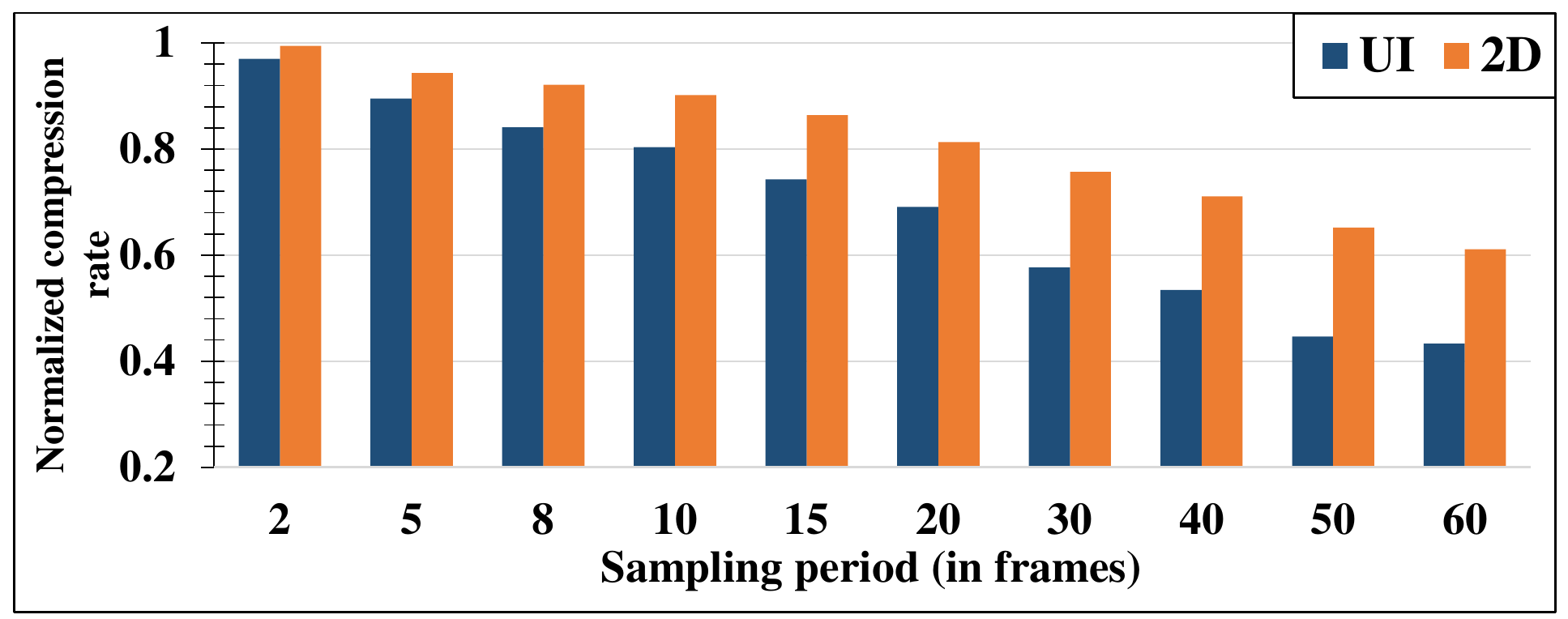} 
	\caption{VDCP normalized (to sampling period of 1) compression rates vs. FVC frame sampling period.} 
	\label{fig:framesampling} 
\end{figure}

\paragraph{\textbf{Frame sampling}}
\label{parag:frame_sampling} 
In frame sampling, the same CCD is used for a number of frames ($N$) instead for just one frame. We vary the sampling period ($N$) for VDCP between 1 (every frame) and 60 frames. Figure~\ref{fig:framesampling} shows compression rates relative to $N$=1. VDCP maintains good compression rates with $N$=2, with a relative compression rate of 97\%. Compression rates, however, significantly decrease with higher $N$ values with 44.6\% and 43.3\% for $N$ values of 50 and 60, respectively. 

\subsection{Implementation Cost and Energy Savings}
\label{sec:hwcost}
\begin{table*}[h] 
	\small 
	\center 
	\resizebox{\textwidth}{!}{
	\begin{tabular}{|c|c|c|c|c|c|c|} 
		\hline 
		\textbf{Structure} & \textbf{Type} & \textbf{Area cost (mm$^2$)} & \textbf{Leakage power (mW)} & \textbf{Access cost (pJ)} & \textbf{Access latency (ns)} & \textbf{Max bandwidth (MPixels/s)} \\ \hline  
		FVC & CAM & 0.00304232 & 0.881899 & 0.766572 & 0.131695 &  7241\\ \hline 
		CCD & CAM & 0.00197988 & 0.39 & 0.402 & 0.128338 & 7431  \\ \hline
		rCCD & Cache & 0.000281592 & 0.281112 & 0.104106 & 0.0722227 & 13203 \\ \hline
	\end{tabular}}
	\caption{DCP structures hardware cost.} 
	\label{table:hwcost} 
\end{table*}

Section~\ref{sec:dcp_schemes} mentions storage requirements associated with DCP. Specifically, for 64-entry FVC/CCD, 456 bytes are need for FVC and 264 bytes for CCD. The cost of rCCDs is (264 bytes) x (maximum number of surfaces that can be read in parallel). Current systems support up to 16 surfaces~\cite{vivantecores}.

For energy, we used DRAMPower v4.0~\cite{chandrasekar2012drampower} to estimate the energy cost of accessing a MICRON 1600\_x32 LPDDR3 DRAM. We found the cost of DRAM accesses to be around 451.2 pJ/byte (this number excludes DRAM idle energy and other system energy costs like the interconnection network). For DCP we used CACTI v7.0~\cite{Balasubramonian2017} with the 22nm process to estimate the area/energy/latency of DCP structures as shown in Table~\ref{table:hwcost}.

Using the numbers in Table~\ref{table:hwcost}, DCP total area cost with support of 16 surfaces equals to 0.009527672 mm$^2$. To compare this area with current hardware, it is less than 0.003\% of Nvidia's Xavier die area~\cite{xavierArea}. For the the dynamic energy cost of compressing/decompression a byte using DCP, we found it to be around 1.3 pJ/byte, i.e., less than 0.29\% of DRAM access cost.

\paragraph{\textbf{Energy savings}} DRAM consumes around 199.6 mW (629.4 mW including static power) for framebuffer operations under a typical rate of 60 FPS using HD frames (GPU writing/display controller reading, or 949.21 MB/s). We calculated DCP total compression/decompression static and dynamic energy consumption (4.83 pJ/byte) and we compared it to only DRAM dynamic energy consumption (451.2 pJ/byte). We found that VDCP reduces the energy consumed by framebuffer operations by 79.9\% for UI apps, 64.4\% for 2D apps, and 41.8\% for 3D apps.

\subsection{Hybrid Schemes} 
\label{subsec:hybrid} 
\begin{figure} 
	\centering 
	\includegraphics[width=.5\textwidth]{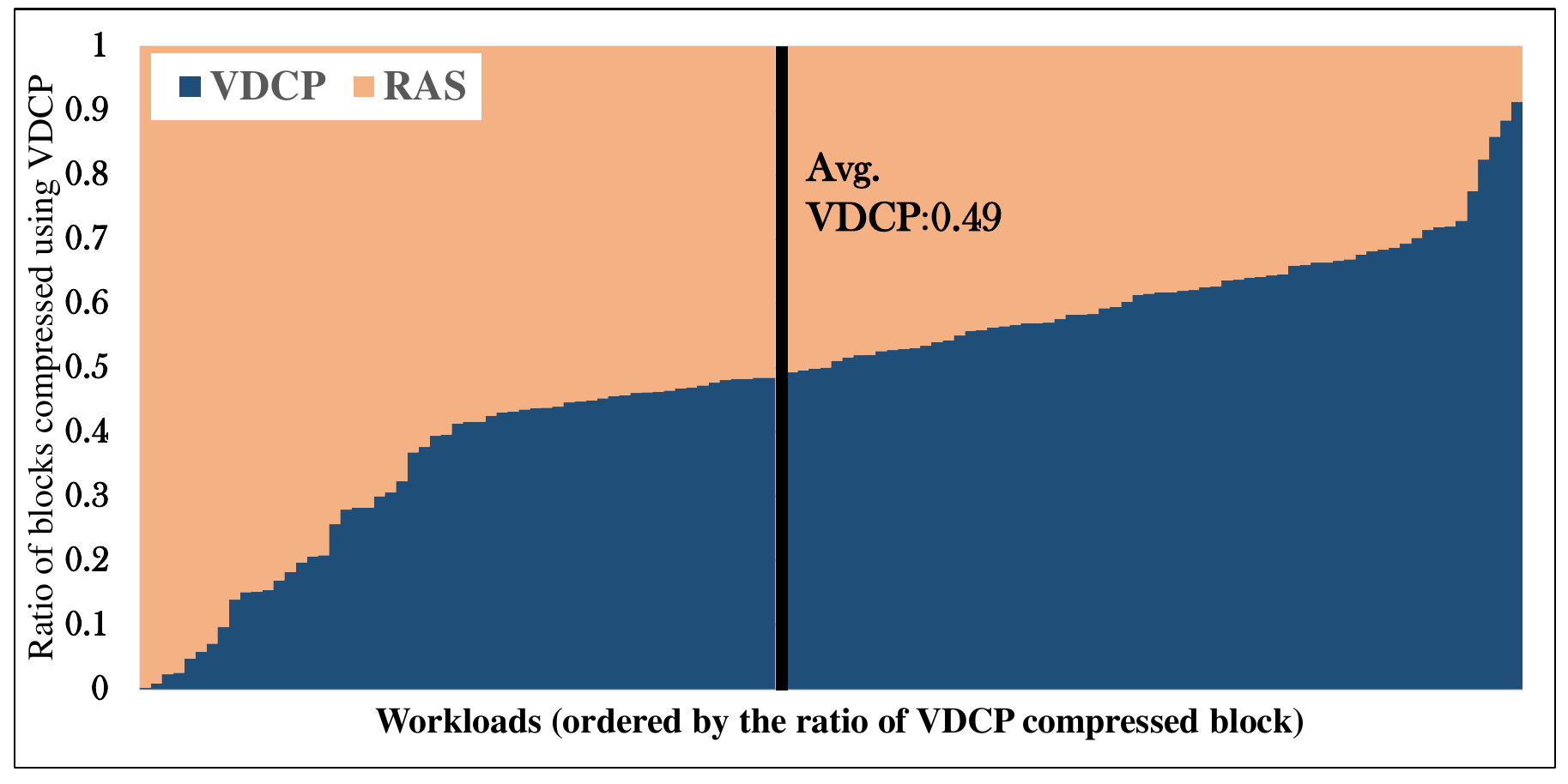} 
	\caption{Ratio of DCP vs. RAS compressed blocks across all workloads.} 
	\label{fig:hybrid} 
\end{figure}

Our hybrid compression scheme uses RAS and VDCP. We compress using both algorithms and then use the best of the two. This exploits VDCP high compression rates for simpler surfaces while falling back on RAS for other cases. RAS+VDCP outperforms RAS and VDCP (with rates of 7.2, 5.206 and 3.23 for UI, 2D and 3D applications respectively). The ratio of VDCP vs. RAS compressed blocks varies by application as shown in Figure~\ref{fig:hybrid}. However, we found that, on average, VDCP and RAS compress an equal number of blocks. 

\section{Conclusion}
\label{sec:conclusion}
This work presents surface compression techniques that reduce the off-chip bandwidth of framebuffer operations in energy-constrained mobile devices. In this work, we analyze and characterize the framebuffer surfaces of UI, 2D and 3D applications and highlight the unique characteristics of each. 

To evaluate our compression schemes, we created and used a set of workloads that represents 124 popular mobile applications. Our results show that VDCP improves compression by an average of 93\% relative to RAS for UI applications, while improving UI and 2D applications over RED by 89\% and 50\%, respectively. 

DCP focuses on 2D and UI applications and can complement other generic compression algorithms. We evaluated a hybrid VDCP+RAS (HDCP) scheme; the scheme was able to increase compression rates by  163\%, 79\% and 27\%  over RAS, and by 159\%, 169\% and 139\% over RED for UI, 2D and 3D applications, respectively.

%

%
\bibliographystyle{IEEEtran}
\bibliography{DCP}

\end{document}